\documentclass[a4paper,oneside,british,english]{aa}
\usepackage[T1]{fontenc}
\usepackage[utf8]{inputenc}
\usepackage{color}
\usepackage{amstext}
\usepackage{graphicx}
\usepackage{subscript}

\makeatletter

\providecommand{\tabularnewline}{\\}
\newcommand{\lyxdot}{.}

\bibpunct{(}{)}{;}{a}{}{,}
\usepackage{babel}

\titlerunning{Birth sites of three young pulsars}

\@ifundefined{showcaptionsetup}{}{%
 \PassOptionsToPackage{caption=false}{subfig}}
\usepackage{subfig}
\makeatother

\usepackage{babel}
\begin{document}

\title{Revisiting the birth locations of pulsars B1929+10, B2020+28,
and B2021+51}

\author{Franz Kirsten\inst{1}\fnmsep\inst{2}\fnmsep\inst{3}\fnmsep\thanks{franz.kirsten@curtin.edu.au, Member of the International Max Planck Research School (IMPRS) for Astronomy and Astrophysics at the Universities of Bonn and Cologne}\and
Wouter Vlemmings\inst{4}\and\foreignlanguage{british}{Robert M. Campbell}\inst{5}\and\foreignlanguage{british}{Michael
Kramer}\inst{2}\and\foreignlanguage{british}{Shami Chatterjee}\inst{6}}

\institute{International Centre for Radio Astronomy Research (ICRAR), Curtin
University, GPO Box U1987, Perth, WA 6845, Australia\and Max Planck
Institut für Radioastronomie (MPIfR), Auf dem Hügel 69, D-53121 Bonn,
Germany\and Argelander Institut für Astronomie (AIfA), Universität
Bonn, Auf dem Hügel 71, D-53121 Bonn, Germany\and Department of Earth
and Space Sciences, Chalmers University of Technology, Onsala Space
Observatory, SE-439 92 Onsala, Sweden\and Joint Institute for VLBI
in Europe, Oude Hoogeveensedijk 4, 7991 PD, Dwingeloo, The Netherlands\and
Department of Astronomy, Cornell University, Ithaca, NY 14853, USA}

\abstract{We present new proper motion and parallax measurements obtained with
the European VLBI Network (EVN) at 5$\,$GHz for the three isolated
pulsars B1929+10, B2020+28, and B2021+51. For B1929+10 we combined our
data with earlier VLBI measurements and confirm the robustness of
the astrometric parameters of this pulsar. For pulsars
B2020+28 and B2021+51 our observations indicate that both stars are
almost a factor of two closer to the solar system than previously
thought, placing them at a distance of $1.39_{-0.06}^{+0.05}$ and
$1.25_{-0.17}^{+0.14}\,$kpc. Using our new astrometry,
we simulated the orbits of all three pulsars in the Galactic potential with the aim
to confirm or reject previously proposed birth locations. Our observations
ultimately rule out a claimed binary origin of B1929+10 and the runaway star
$\zeta$ Ophiuchi in Upper Scorpius. A putative common binary origin
of B2020+28 and B2021+51 in the Cygnus Superbubble is also very unlikely.}


\maketitle

\section{Introduction }

\begin{table*}
\begin{centering}
\caption[\selectlanguage{british}%
Phase calibrator details for observations of pulsars B1929+10, B2020+28,
B2021+51\selectlanguage{english}%
]{\label{tab:observations-details}\foreignlanguage{british}{}Observing
epochs and arrays.}
\begin{tabular}{lcccccccccccccc}
\noalign{\vskip-0.3cm}
\hline\hline &  &  &  &  &  &  &  &  &  &  &  &  &  & \tabularnewline
Date(s) & UT range & \multicolumn{13}{l}{Array\tablefootmark{a}\textsuperscript{}}\tabularnewline
\hline 
\noalign{\vskip\doublerulesep}
2010 May 28-29 & 2130--0930 & Ef & Wb & Jb & Jv & On & Mc & Tr &  &  & Zc & Bd & Ur & Sh\tabularnewline
2010 Oct 27 & 1130--2330 & Ef & Wb &  &  & On & Mc & Tr & Ys & Sv & Zc & Bd &  & \tabularnewline
2011 Mar 09 & 0300--1500 & Ef & Wb & Jb & Jv & On & Mc & Tr & Ys & Sv & Zc & Bd & Ur & Sh\tabularnewline
2011 Jun 01-02 & 2120--0920 & Ef & Wb & Jb &  & On &  & Tr & Ys & Sv & Zc & Bd & Ur & Sh\tabularnewline
\hline\vspace{-0.4cm} &  &  &  &  &  &  &  &  &  &  &  &  &  & \tabularnewline
\end{tabular}
\par\end{centering}

\tablefoot{\tablefoottext{a}{Ef = Effelsberg, DE, $100\,$m; Wb = Westerbork Synthesis Radio Telescope, NL, $12-14\times25\,$m; Jb = Jodrell Bank Lovell Telescope, UK, $76\,$m; Jv = Jodrell Bank Mark2 Telescope, UK, $25\,$m; On = Onsala, SE, $25\,$m; Mc = Medicina, IT, $32\,$m; $\text{Tr}=\text{Torun}$, PL, $32\,$m; Ys = Yebes, ES, $40\,$m; Sv = Svetloe, RU, $32\,$m; Zc = Zelenchukskaya, RU, $32\,$m; Bd = Badary, RU, $32\,$m; Ur = Urumqi, CN, $32\,$m; Sh = Shanghai, CN, $25\,$m.}} 
\end{table*}
\begin{table*}
\centering{}\caption[\selectlanguage{british}%
Phase calibrator details for observations of pulsars B1929+10, B2020+28,
B2021+51\selectlanguage{english}%
]{\label{tab:Calibrator-details}\foreignlanguage{british}{}Calibrator
details for each pulsar.}
\begin{tabular}{lcccc}
\noalign{\vskip-0.3cm}
\hline\hline &  &  &  & \tabularnewline
\noalign{\vskip-0.2cm}
 & \multicolumn{2}{c}{Pointing centre} & Distance to & Flux density\tablefootmark{b}\tabularnewline
Source & RA & Dec & pulsar {[}deg{]} & {[}mJy$\,$beam$^{-1}${]}\tabularnewline
\hline 
\noalign{\vskip\doublerulesep}
B1929+10 & 19:32:14.0160 & 10:59:32.868 &  & 50\tabularnewline
\hspace{0.2cm}J1928+0848\tablefootmark{a} & 19:28:40.8555 & 08:48:48.413 & 2.35 & 160\tabularnewline
\hspace{0.2cm}J1934+1043 & 19:34:35.0256 & 10:43:40.366 & 0.63 & 50\tabularnewline
 &  &  &  & \tabularnewline[-0.25cm]
B2020+28 & 20:22:37.0697 & 28:54:22.976 &  & 30\tabularnewline
\hspace{0.2cm}J2020+2826\tablefootmark{a} & 20:20:45.8707 & 28:26:59.195 & 0.61 & 70\tabularnewline
\hspace{0.2cm}J2023+3153 & 20:23:19.0173 & 31:53:02.306 & 2.98 & 900\tabularnewline
 &  &  &  & \tabularnewline[-0.25cm]
B2021+51 & 20:22:49.8596 & 51:54:50.400 &  & 80\tabularnewline
\hspace{0.2cm}J2025+5028\tablefootmark{a} & 20:25:24.9725 & 50:28:39.536 & 1.49 & 110\tabularnewline
\hspace{0.2cm}J2023+5427 & 20:23:55.8440 & 54:27:35.829 & 2.55 & 500\tabularnewline
\hline\vspace{-0.4cm} &  &  &  & \tabularnewline
\end{tabular} \tablefoot{
\tablefoottext{a}{These are the calibrators referred to as primary calibrators in the text.}
\tablefoottext{b}{For the pulsars this is the apparent pulsar flux density measured by employing pulse gating.}
}
\end{table*}
Typical transverse velocities of isolated pulsars are of the order
of several hundred km$\,$s$^{-1}$ \citep{cordes98,arzoumanian02,hobbs05},
while those of their progenitor O- and B-stars are at most several
tens of km$\,$s$^{-1}$. In the standard neutron star formation scenario
this discrepancy is explained by an asymmetry in the supernova (SN)
explosion that imparts a kick to the forming central compact object,
accelerating it to the observed velocities \citep[e.g.][]{scheck06}.
As a result of the short lifetime of SN remnants ($<10^{5}\,$yr) and the
typical characteristic age of young pulsars ($\tau_{c}\sim10^{6-7}\,$Myr),
direct associations between SN-remnants and pulsars are rare. Measurements
of accurate proper motions $\mu$ and parallaxes $\pi$ of pulsars
can, however, indicate the birth locations of pulsars. The combination
of both $\mu$ and the distance $d=1/\pi$ yields the physical transverse
velocity, $V_{\perp}$, which, given an estimate of the radial velocity,
$V_{r}$, allows calculating a trajectory that traces the pulsar
back to its possible birth location. Hence, kinematic ages -- as opposed
to characteristic ages $\tau_{c}=P/2\dot{P}$ -- of pulsars can be
determined and conclusions about neutron star formation scenarios
can be drawn. 

One of the first to calculate pulsar orbits was \citet{wright79},
claiming that the pulsars B1929+10 (J1932+1059) and B1952+29 originate
from a former binary system. More recently, \citet{hoogerwerf01}
used the 3D space velocity of high-velocity runaway stars and parallax
and proper motion measurements of young nearby pulsars to extrapolate
their trajectories back in time. Their simulations indicated that
the runaway O-star $\zeta$ Ophiuchi ($\zeta$ Oph, HIP 81377) and
the young pulsar B1929+10 were likely to have been in a binary system
in Upper Scorpius (Scorpius-Centaurus association) until about 1$\,$Myr
ago. According to their analysis, the system was disrupted when the
progenitor star underwent a supernova explosion. During that event,
the space velocity vectors of both $\zeta$ Oph and the pulsar were
modified to point away from Upper Scorpius. The parameter range for
which such a scenario is possible is, however, rather small. Improved
measurements of $\mu$ and $\pi$ for B1929+10 obtained with the NRAO
Very Long Baseline Array (VLBA) led to the conclusion that a common
origin of the pulsar and $\zeta$ Oph is unlikely \citep{chatterjee04}.
Adopting the measurements of \citet{chatterjee04}, but increasing
the reported uncertainties by factors between 10 and 30, \citet{bobylev08}
and also \citet{tetzlaff10} repeated the simulations of \citet{hoogerwerf01},
re-postulating a binary origin of B1929+10 and $\zeta$ Oph in Upper
Scorpius.

In a similar investigation, \citet{vlemmings04} identified the two
pulsars B2020+28 and B2021+51 as candidates for a common-origin scenario
based on proper motion and parallax measurements obtained with the
VLBA at 1.4$\,$GHz \citep{brisken02}. The authors simulated the
trajectories of the two pulsars back in time and concluded that they
most likely originated from a binary system in the Cygnus Superbubble that
was disrupted when the younger of the two pulsars was born in a supernova.
The measurements by \citet{brisken02} are, however, based on five
observations covering a time span of only roughly one year. 

Here, we present new measurements of $\mu$ and $\pi$ for the three
pulsars B1929+10, B2020+28, and B2021+51 obtained with the European
VLBI Network (EVN) at an observing frequency of 5$\,$GHz. These observations
extend the time baseline to more than ten years, allowing for an extremely
high precision in measurements of $\mu$ and $\pi$ for all three
pulsars. We used these data and ran new simulations of trajectories to shed new light on the proposed binary origin of B1929+10/$\zeta$ Oph and B2020+28/B2021+51.

\section{Observations and data reduction\label{sec:Observations-and-reduction}}

\begin{figure}
\begin{centering}
\includegraphics[width=1\columnwidth]{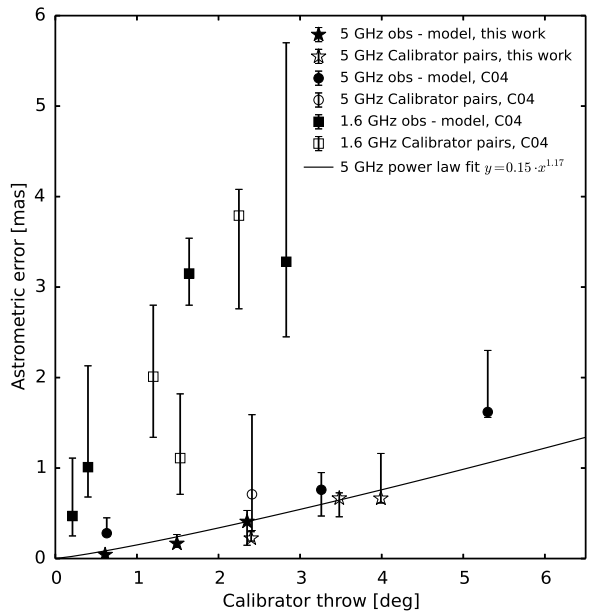}
\par\end{centering}

\caption[Astrometric accuracy as a function of angular separation between phase
calibrator source and target]{\label{fig:cal-throw}Astrometric accuracy as a function of angular
separation between phase calibrator source and target. This is a reproduction
of Fig. 3 in \citet{chatterjee04} (referred to as C04 here) to which
we added data from our observations (open and filled stars). Open
symbols denote the median scatter about the average position of the
primary phase calibrator obtained from calibrating its visibilities
with solutions from the secondary calibrator. Filled symbols show
the median scatter of the observed positions about the best-fit model
for proper motion and parallax. The 1.6$\,$GHz data (squares) are
taken from \citet{chatterjee01} and \citet{vlemmings03}, while the previous
5$\,$GHz data (circles) are taken from \citet{chatterjee04}. The solid
line is a weighted least-squares power-law fit to all 5$\,$GHz data
points. The almost linear relationship between astrometric error and
calibrator throw seems to break down beyond an angular separation
of more than about four degrees.}
\end{figure}
The observations described here were conducted with the EVN under
project code EV018(A-D). All observations used a frequency range $4926.49-5054.49\,$MHz
with dual circular polarizations and two-bit sampling, for a total bit-rate
of 1 Gbps per station. The 128$\,$MHz frequency range in each polarization
was split into eight 16$\,$MHz baseband channels. We conducted four
epochs of observations between May 2010 and June 2011, as summarized
in Table \ref{tab:observations-details}. This table also lists the
EVN stations that successfully participated in the array at each epoch.
\\
\\
We observed each of the three pulsars in all four 12-hour epochs,
using phase-referencing. Table \ref{tab:Calibrator-details} summarizes
the pulsars and characteristics of their phase-reference calibrator
sources. Our phase-referencing tactics included (i) a basic four-minute
cycle alternating between the target pulsar (2.5$\,$min) and the
primary calibrator (1.5$\,$min), and (ii) insertion of an additional
1.5 minute scan of the secondary calibrator in every second cycle.
For bandpass calibration, we observed the quasars J1800+3848 and 3C454.3:
the former about two hours from the beginning and the latter four
hours from the end of each epoch. This observing pattern provides
about two hours of integration on each pulsar per epoch, yielding
a nominally expected sensitivity of the arrays ranging from 10.9 to
14.8$\,\mu$Jy per beam.\\
\\
The data from the telescopes were correlated on the EVN software correlator
at the Joint Institute for VLBI in Europe (SFXC; \citealt[]{keimpema15}).
Each of the 16 MHz baseband channels was correlated with 32 frequency
points and  one-second coherent integrations. Pulsar scans were correlated
using the gating/binning capability of SFXC. Because these were the
first observations to employ this mode, we devote a few sentences
to describe it. Given an ephemeris of a pulsar, SFXC can apply a gate,
defined by a start and stop fraction of a period, such that correlation
accumulates only during the in-gate interval. Before gating, the
pulsar data are de-dispersed. Here, we used incoherent de-dispersion
(a constant correction per $0.5\sim\,$MHz frequency point). Coherent
de-dispersion has subsequently been developed on SFXC. The gate itself
may be divided further into a number of equal-width bins, each of
which produces independent correlator output. In this case, there
was only one bin. In this way, assuming the entire pulse falls within
the gate, the signal-to-noise ratio (S/N) of the pulsar detections
can be increased by a factor of about $\sqrt{P/w}$, where $P$ is
the pulse period and $w$ is the width of the gate. The pulsar ephemerides
were derived with TEMPO2%
\footnote{http://www.atnf.csiro.au/research/pulsar/tempo2/%
} \citep{hobbs06}. Before full correlation, we conducted iterations
of gate-fitting, using a full-period gate with 40 bins, to confirm
that the pulse profile was stationary over the time-range of a pulsar
observation within an epoch and to optimize the choice of the gate
start/stop parameters. For these three pulsars, the gate widths used
were typically in the range of $4-8\%$ of a pulse period, leading
to gating gain factors of $\sim3-5$.\\
\\
We performed a mostly automated data reduction and calibration procedure
relying on the NRAO Astronomical Image Processing System (AIPS) and
the scripting language ParselTongue \citep{kettenis06}. Removal of
data affected by radio frequency interference (RFI) was made running
the RFI-mitigation software SERPent \citep{peck13} and some further
manual flagging. We first applied the system temperature and gain
curve corrections as determined by the EVN pipeline%
\footnote{http://www.evlbi.org/pipeline/user\_expts.html%
} and also corrected for the parallactic angle using the AIPS task
CLCOR. Next we computed ionospheric corrections in TECOR with the
help of total electron content maps as published by the Center for
Orbit Determination in Europe%
\footnote{ftp://ftp.unibe.ch/aiub/CODE/%
}. In a first calibration run, we solved for visibility rates, phases,
and delays in FRING for all calibrator sources assuming a simple point
source model. Next, we self-calibrated on each source, improving the
S/N by a factor of five to ten. For each calibrator we then concatenated
the calibrated data from all four epochs. This dataset was imaged
to produce a global model of each calibrator source. The dominant
CLEAN components of each source model were then used as the input
model parameters in a second FRING-run. In this way, we eliminated
any systematics
caused by source structure that affected the position of the calibrator
sources in between epochs. For each of the three pulsars
the calibration solutions of the primary calibrator were of much higher
quality than those of the secondary calibrator and, hence, were applied
to the respective target pulsar (Table \ref{tab:Calibrator-details}).

We used the calibration solutions from the secondary calibrator to
provide independent checks on the achieved astrometric accuracy as
a function of angular separation between target and calibrator source
(`calibrator throw'), as done by \citet{chatterjee04}. Similarly to
the earlier data, our observations imply an almost linear increase
in astrometric accuracy with decreasing calibrator throw (Fig. \ref{fig:cal-throw}).
We fitted a power law, $y=a\cdot x^{b}$, to all data points obtained
from 5$\,$GHz observations, which yielded $(a,\, b)=(0.15\pm0.09,\,1.17\pm0.31)$.
This relation, however, seems to only hold for angular separations
of up to four degrees between calibrator and target.

\section{Estimates for astrometric parameters of B1929+10, B2020+28, and B2021+51\label{sub :Estimating-astrometric-parameters}}

\begin{figure*}
\begin{centering}
\includegraphics[width=0.25\textwidth]{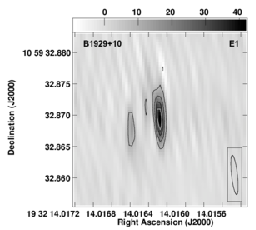}\includegraphics[width=0.25\textwidth]{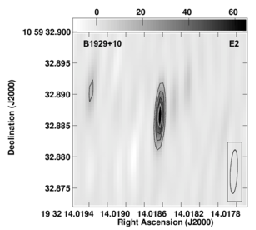}\includegraphics[width=0.25\textwidth]{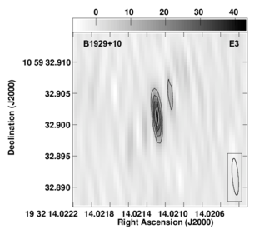}\includegraphics[width=0.25\textwidth]{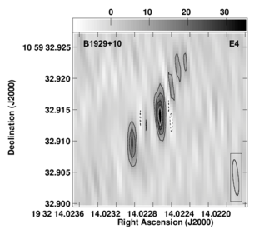}
\par\end{centering}

\begin{centering}
\includegraphics[width=0.25\textwidth]{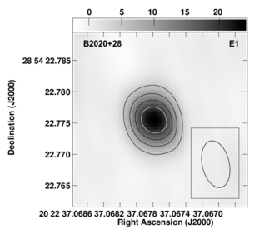}\includegraphics[width=0.25\textwidth]{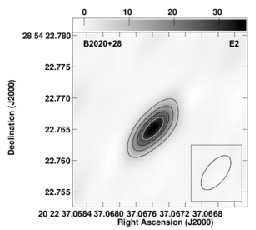}\includegraphics[width=0.25\textwidth]{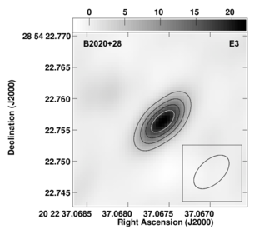}\includegraphics[width=0.25\textwidth]{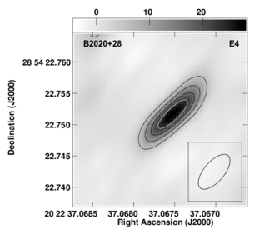}
\par\end{centering}

\begin{centering}
\includegraphics[width=0.25\textwidth]{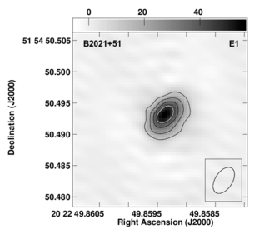}\includegraphics[width=0.25\textwidth]{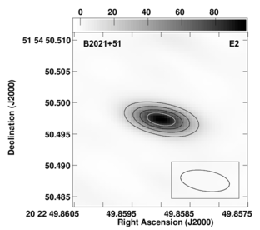}\includegraphics[width=0.25\textwidth]{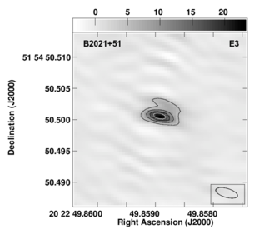}\includegraphics[width=0.25\textwidth]{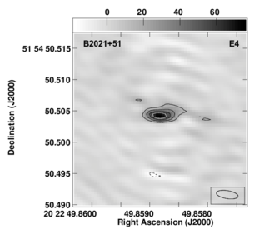}
\par\end{centering}

\caption{\label{fig:pulsar-kntr-plots}Greyscale plots of the fitted pulsar
images. Rows from top to bottom show B1929+10, B2020+28, and
B2021+51. Columns from left to right correspond to epochs one
to four. Overlaid contours increase in steps of 20 percent of the
peak flux density, where negative values are indicated by dashed contours.
The absolute flux density scale (mJy$\,$beam$^{-1}$) is indicated
above each individual panel, and the beam size and position angle are
indicated in the bottom right. The larger beam sizes for B2020+28
and B2021+51 compared to that of B1929+10 are due to flagging of baselines
to Sh and/or Ur.}
\end{figure*}
\begin{table*}[t]
\caption[\selectlanguage{british}%
Measured positions of B1929+10, B2020+28, B2021+51 at MJD 55630\selectlanguage{english}%
]{\label{tab:Measured-psr-positions}\foreignlanguage{british}{}Measured
positions at MJD 55629. }

\begin{centering}
\begin{tabular}{ccccc}
\noalign{\vskip-0.3cm}
\hline\hline &  &  &  & \tabularnewline
\noalign{\vskip-0.2cm}
Pulsar & RA (J2000) & Dec (J2000) & S/N & beam size {[}mas$\times$mas{]}\tabularnewline
\hline 
\noalign{\vskip\doublerulesep}
B1929+10 & 19:32:14.021289(1) & 10:59:32.90137(5) & 155 & $1.01\times5.94$\tabularnewline
B2020+28 & 20:22:37.06758(1) & 28:54:22.7563(2) & 142 & $3.76\times7.36$\tabularnewline
B2021+51 & 20:22:49.85890(1) & 51:54:50.5005(1) & 156 & $1.72\times3.97$\tabularnewline
\hline\vspace{-0.4cm} &  &  &  & \tabularnewline
\end{tabular}
\par\end{centering}

\tablefoot{Numbers in brackets indicate the uncertainty in the last digit. The larger beam size for B2020+28 is due to flagging of baselines to Sh and Ur.}
\end{table*}
\begin{table*}
\caption[\selectlanguage{british}%
Previous proper motion and parallax estimates for pulsars B1929+10,
B2020+28, B2021+51\selectlanguage{english}%
]{\label{tab:previous-fitting-results}\foreignlanguage{british}{}Previous
proper motion and parallax estimates and derived values.}

\begin{centering}
\begin{tabular}{ccccccc}
\noalign{\vskip-0.3cm}
\hline\hline &  &  &  &  &  & \tabularnewline
\noalign{\vskip-0.2cm}
 & $\mu_{\alpha}$ & $\mu_{\delta}$ & $\pi$ & \selectlanguage{british}%
$d$\selectlanguage{english}%
 & \selectlanguage{british}%
$V_{\perp}$\selectlanguage{english}%
 & \tabularnewline
Pulsar & {[}mas$\,$yr$^{-1}${]} & {[}mas$\,$yr$^{-1}${]} & {[}mas{]} & \selectlanguage{british}%
{[}kpc{]}\selectlanguage{english}%
 & \selectlanguage{british}%
{[}km$\,$s$^{-1}${]}\selectlanguage{english}%
 & Reference\tabularnewline
\hline 
\noalign{\vskip\doublerulesep}
B1929+10 & $99.0\pm12.0$ & $39.0\pm8.0$ & $4.00\pm2.0$ & \selectlanguage{british}%
$0.24_{-0.12}^{+0.09}$\selectlanguage{english}%
 & $124_{-52}^{+140}$ & 1\tabularnewline
B1929+10 & $94.82\pm0.26$ & $43.04\pm0.15$ & $3.02\pm0.09$ & \selectlanguage{british}%
$0.33_{-0.01}^{+0.01}$\selectlanguage{english}%
 & $162_{-5}^{+6}$ & 2\tabularnewline
\noalign{\vskip\doublerulesep}
B1929+10 & $94.09\pm0.11$ & $42.99\pm0.16$ & $2.77\pm0.07$ & \selectlanguage{british}%
$0.361_{-0.009}^{+0.009}$\selectlanguage{english}%
 & $176_{-5}^{+5}$ & 3\tabularnewline
B2020+28 & $-4.38\pm0.53$ & $-23.59\pm0.26$ & $0.37\pm0.12$ & \selectlanguage{british}%
$2.70_{-0.96}^{+0.64}$\selectlanguage{english}%
 & $256_{-62}^{+135}$ & 2\tabularnewline
B2021+51 & $-5.23\pm0.17$ & $11.54\pm0.28$ & $0.50\pm0.07$ & \selectlanguage{british}%
$2.00_{-0.31}^{+0.23}$\selectlanguage{english}%
 & $149_{-22}^{+28}$ & 2\tabularnewline
\hline\vspace{-0.4cm} &  &  &  &  &  & \tabularnewline
\end{tabular}
\par\end{centering}

\centering{}\tablebib{(1) \citet{hoogerwerf01}; (2) \citet{brisken02}; (3) \citet{chatterjee04}}
\end{table*}
\begin{table*}
\caption[Astrometric parameters and derived values for pulsars B1929+10, B2020+28,
and B2021+51]{\label{tab:my-fitting-results}Astrometric results and derived values
from the estimation strategies.}

\begin{centering}
\begin{tabular}{cccccccccc}
\noalign{\vskip-0.3cm}
\hline\hline &  &  &  &  &  &  &  &  & \tabularnewline
\noalign{\vskip-0.2cm}
 & \multicolumn{2}{c}{data sets\tablefootmark{a}} &  & $\mu_{\alpha}$ & $\mu_{\delta}$ & $\pi$ &  & $d$ & $V_{\perp}$\tabularnewline
Pulsar & F\tablefootmark{b} & B\tablefootmark{c} & N$_{{\rm {obs}}}$ & {[}mas$\,$yr$^{-1}${]} & {[}mas$\,$yr$^{-1}${]} & {[}mas{]} & $\chi_{\text{red}}^{2}$ & {[}kpc{]} & {[}km$\,$s$^{-1}${]}\tabularnewline
\hline 
\noalign{\vskip\doublerulesep}
B1929+10 & C &  & $4$ & $94.11\pm0.52$ & $43.64\pm1.06$ & $2.79\pm0.14$ & $1.34$ & $0.358_{-0.019}^{+0.017}$ & $175_{-10}^{+11}$\tabularnewline[0.2cm]
\noalign{\vskip\doublerulesep}
 & C\textsubscript{C}, C &  & $10$ & $94.04\pm0.12$ & $43.39\pm0.23$ & $2.79\pm0.08$ & $1.03$ & $0.358_{-0.010}^{+0.010}$ & $175_{-5}^{+5}$\tabularnewline[0.2cm]
\noalign{\vskip\doublerulesep}
 & L, C &  & $9$ & $94.20\pm0.20$ & $42.93\pm0.26$ & $2.78\pm0.05$ & $0.51$ & $0.360_{-0.006}^{+0.006}$ & $175_{-3}^{+4}$\tabularnewline[0.2cm]
\noalign{\vskip\doublerulesep}
 & L, C\textsubscript{C}, C &  & $15$ & $94.06\pm0.09$ & $43.24\pm0.17$ & $2.78\pm0.06$ & $0.77$ & $0.360_{-0.008}^{+0.007}$ & $176_{-5}^{+4}$\tabularnewline[0.2cm]
\noalign{\vskip\doublerulesep}
 &  & C\textsubscript{C}, C & $10$ & $94.07_{-0.20}^{+0.14}$ & $43.41_{-0.12}^{+0.10}$ & $2.76_{-0.14}^{+0.10}$ &  & $0.362_{-0.012}^{+0.019}$ & $177_{-7}^{+7}$\tabularnewline[0.2cm]
\noalign{\vskip\doublerulesep}
 &  & L, C & $9$ & $94.23{}_{-0.24}^{+0.14}$ & $42.95{}_{-0.26}^{+0.23}$ & $2.79_{-0.12}^{+0.07}$ &  & $0.362_{-0.012}^{+0.019}$ & $175_{-5}^{+8}$\tabularnewline[0.2cm]
\noalign{\vskip\doublerulesep}
 &  & L, C\textsubscript{C}, C & $15$ & $94.08_{-0.17}^{+0.13}$ & $43.25{}_{-0.13}^{+0.16}$ & $2.77_{-0.07}^{+0.08}$ &  & $0.361_{-0.010}^{+0.009}$ & $176_{-5}^{+5}$\tabularnewline[0.2cm]
\hline &  &  &  &  &  &  &  &  & \tabularnewline[-0.2cm]
B2020+28 & C &  & $4$ & $-3.34\pm0.05$ & $-23.65\pm0.11$ & $0.72\pm0.03$ & $0.05$ & $1.39_{-0.06}^{+0.05}$ & $134{}_{-6}^{+6}$\tabularnewline[0.2cm]
\noalign{\vskip\doublerulesep}
 & L, C &  & $9$ & $-3.46\pm0.17$ & $-23.73\pm0.21$ & $0.61\pm0.08$ & $0.53$ & $1.63_{-0.23}^{+0.18}$ & $158_{-20}^{+25}$\tabularnewline[0.2cm]
\noalign{\vskip\doublerulesep}
 &  & L, C & $9$ & $-3.45{}_{-0.33}^{+0.16}$ & $-23.70{}_{-0.22}^{+0.32}$ & $0.60{}_{-0.14}^{+0.13}$ &  & $1.66_{-0.29}^{+0.50}$ & $160_{-30}^{+50}$\tabularnewline[0.2cm]
\hline &  &  &  &  &  &  &  &  & \tabularnewline[-0.2cm]
B2021+51 & C &  & $4$ & $-5.08\pm0.42$ & $10.84\pm0.25$ & $0.80\pm0.11$ & $1.54$ & $1.25_{-0.17}^{+0.14}$ & $87{}_{-12}^{+15}$\tabularnewline[0.2cm]
\noalign{\vskip\doublerulesep}
 & L, C &  & $9$ & $-5.03\pm0.27$ & $10.96\pm0.17$ & $0.78\pm0.07$ & $0.90$ & $1.28_{-0.12}^{+0.10}$ & $90_{-9}^{+11}$\tabularnewline[0.2cm]
\noalign{\vskip\doublerulesep}
 &  & L, C & $9$ & $-5.01{}_{-0.20}^{+0.17}$ & $10.99{}_{-0.29}^{+0.18}$ & $0.77{}_{-0.10}^{+0.08}$ &  & $1.30_{-0.13}^{+0.19}$ & $91_{-12}^{+16}$\tabularnewline
\hline\vspace{-0.4cm} &  &  &  &  &  &  &  &  & \tabularnewline
\end{tabular}
\par\end{centering}

\tablefoot{
\tablefoottext{a}{C refers to the measurements obtained in this campaign, C$_\text{C}$ denotes the data set from Chatterjee et al. (2004), and L indicates that the measurements from Brisken et al. (2002) were included in the analysis.}
\tablefoottext{b}{Results derived from a least-squares fit of the measured data.}
\tablefoottext{c}{Median values and $68\%$ confidence interval from fitting $10^5$ bootstrapped realizations of the data.}
}
\end{table*}
\begin{figure*}
\includegraphics[width=1\textwidth]{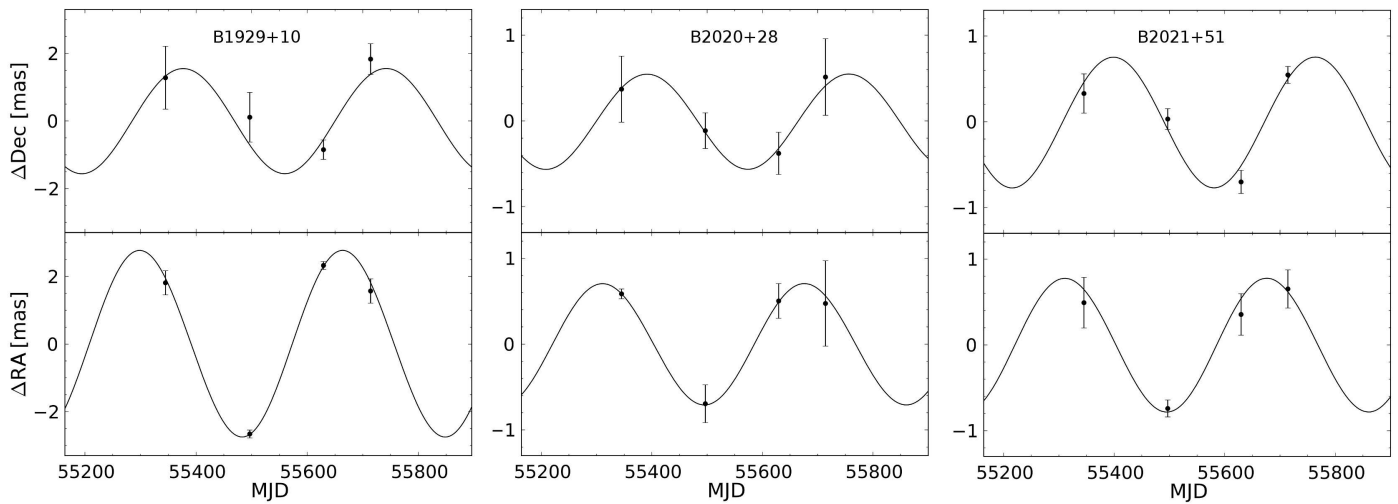}

\caption[Parallax signatures of pulsars B1929+10, B2020+28, and B2021+51 from
5$\,$GHz data only]{\label{fig: c-band-fits}Relative measured positions of B1929+10
(left), B2020+28 (middle), and B2021+51 (right) with the best-fit
proper motion removed. The solid line is the best-fit parallax from
our EVN 5$\,$GHz observations.}
\end{figure*}
\begin{figure*}
\subfloat{\begin{centering}
\includegraphics[width=0.33\textwidth]{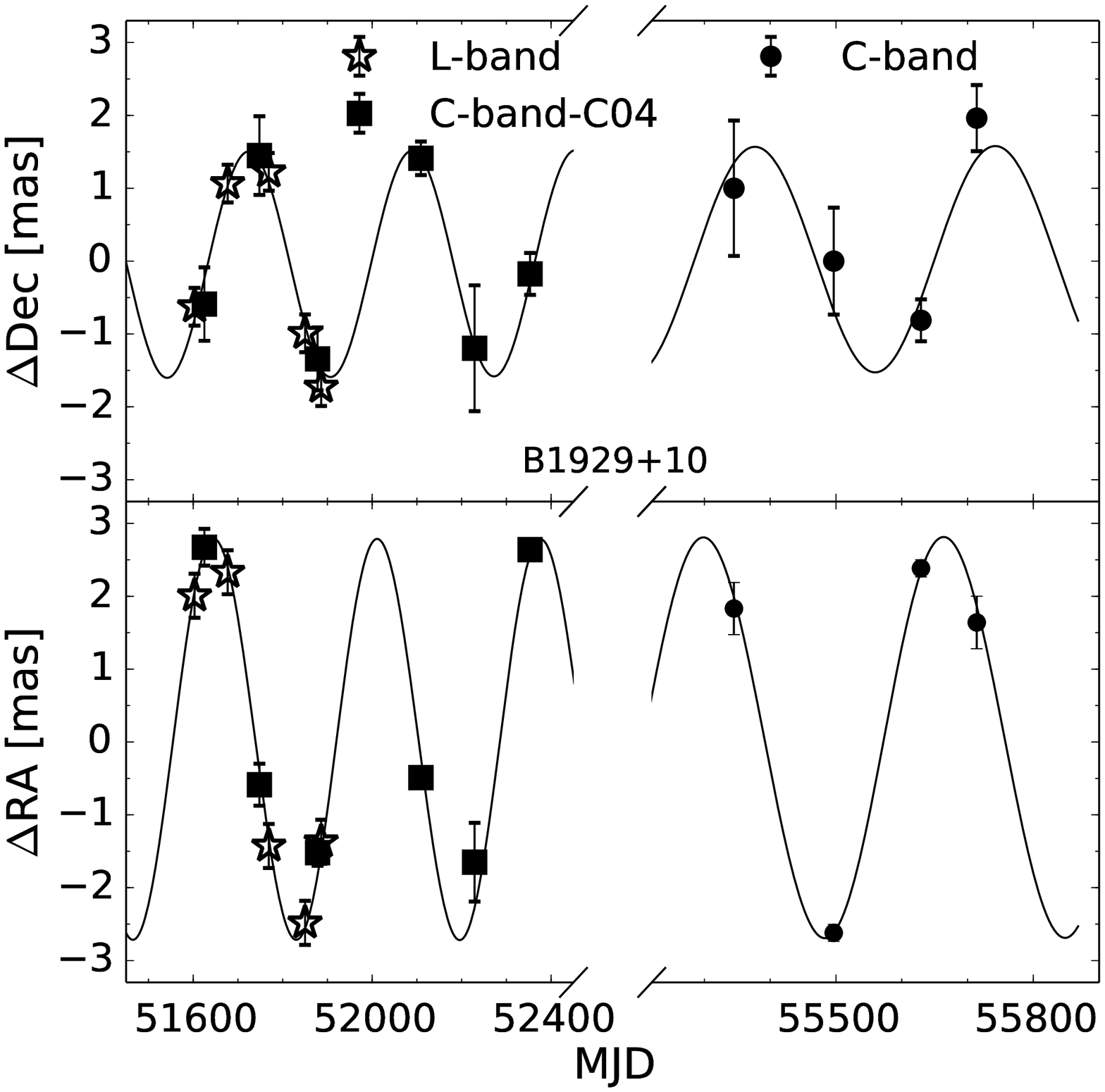}\includegraphics[width=0.33\textwidth]{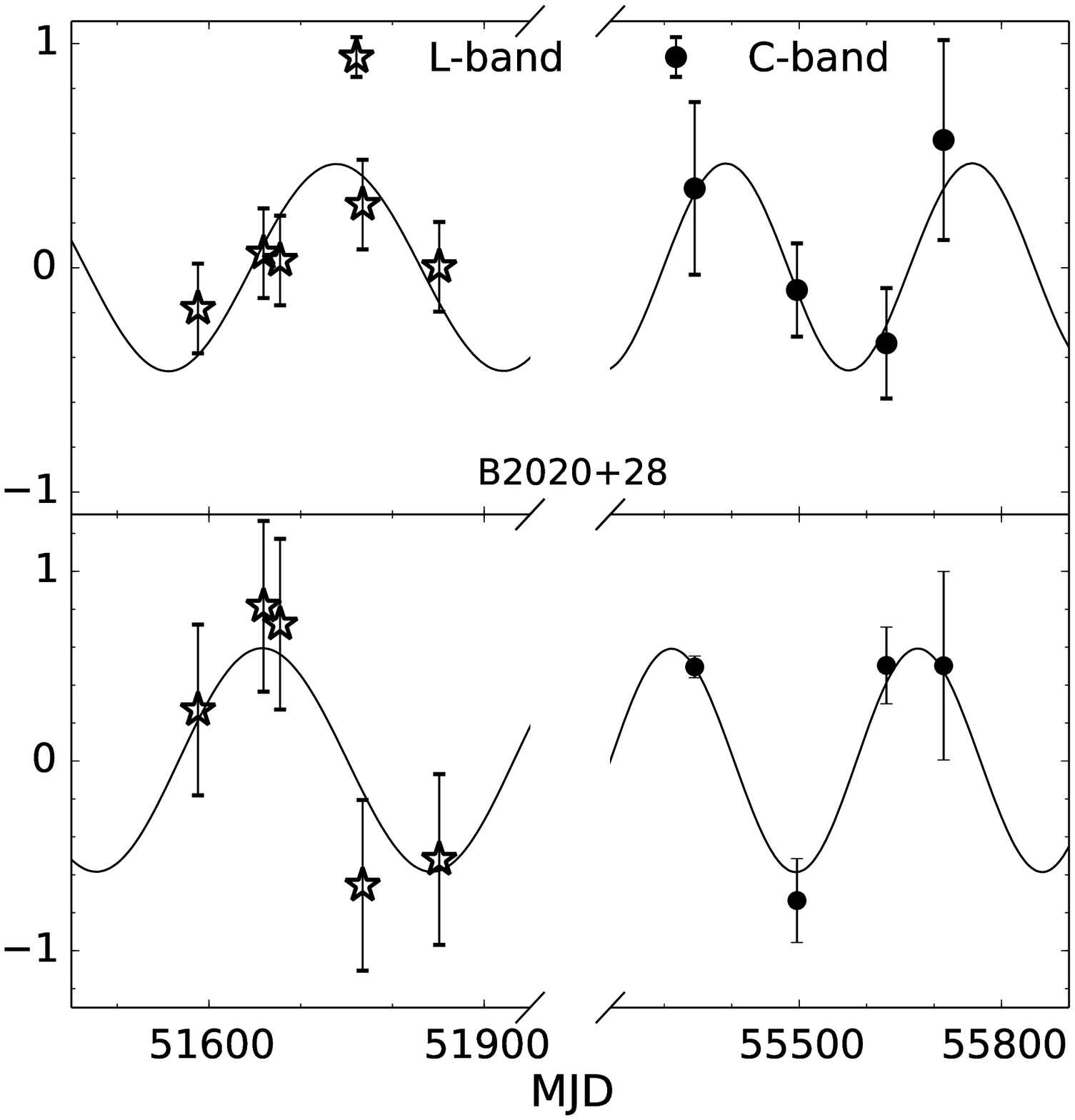}\includegraphics[width=0.33\textwidth]{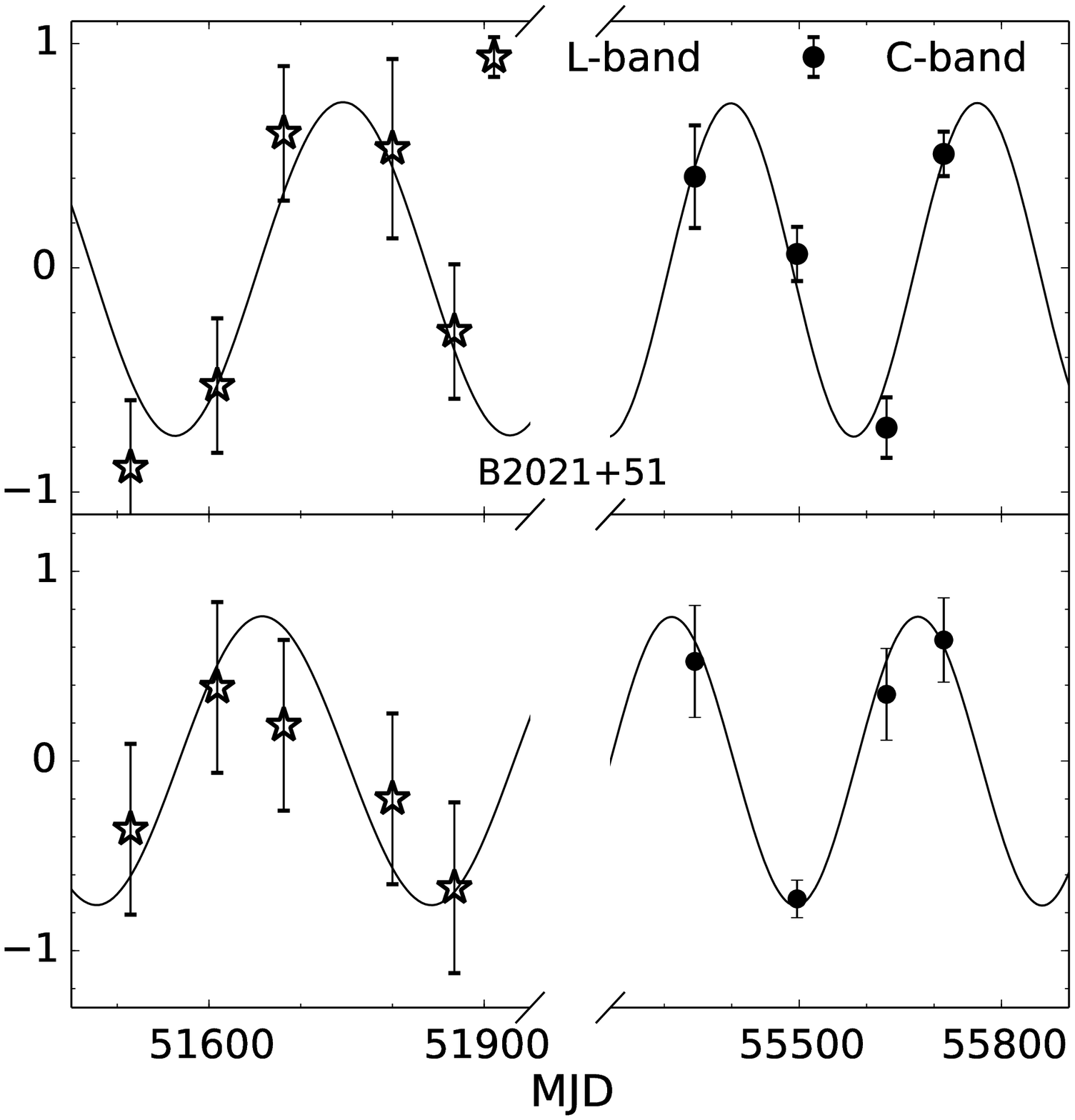}
\par\end{centering}

}

\caption[Parallax signatures of pulsars B1929+10, B2020+28, and B2021+51 from
all available data.]{\label{fig: all-bands-fits}Same as Fig. \ref{fig: c-band-fits},
but for the fits performed using all available data. Solid dots show
the most recent measurements at 5 GHz, open stars are the data taken from
\citet{brisken02}, and squares are the data adopted from \citet{chatterjee04}.
For better illustration, we omitted the time span without any observations.}
\end{figure*}
We measured the position of the pulsars in each epoch by fitting a
2D Gaussian to the brightness distribution in the image plane (Fig.
\ref{fig:pulsar-kntr-plots}) using the AIPS task IMFIT. As a
result of the
high S/N ($\sim150$) and small beam size ($\theta\sim3\times5\,$mas),
the formal errors are very small ($\theta/(2*\text{{S/N}})\sim10\times20\,\mu$as),
certainly underestimating the real positional uncertainties. In addition
to these random errors, residual systematic errors caused by
the calibrator throw (Fig. \ref{fig:cal-throw}), for instance,
 need to be taken
into account. A good estimate for these systematic errors is the deconvolved
size $\theta_{\text{d}}$ of the pulsar, which is zero for a true
point source. Following the scheme reported in \citet{chatterjee01}, we estimated
the systematic uncertainties using the quantity $\theta_{\text{d}}/\sqrt{{(N_{\text{ant}}-1)*t_{\text{obs}}/t_{\text{iono}}}}$,
where $N_{\text{ant}}=11$ is the typical number of antennas involved,
$t_{\text{obs}}=110\,$min is the total amount of time spent on each
pulsar, and $t_{\text{iono}}=6\,$min is the empirically determined
atmospheric coherence time at 5$\,$GHz. For the total positional
uncertainty we added both the formal and the systematic errors in quadrature.
Table \ref{tab:Measured-psr-positions} lists the measured positions
of all three pulsars in the third epoch at MJD 55629. \\
\\
To estimate each pulsar's proper motion and parallax, we performed a
weighted least-squares-fit to the measured positions. Here, we measured
both parameters in three ways: we considered our position measurements
alone (Fig. \ref{fig: c-band-fits}), we combined our data with those
of the publications listed in Table \ref{tab:previous-fitting-results}
(Fig. \ref{fig: all-bands-fits}), and we employed a bootstrapping
technique. For the latter, we randomly sampled the position measurements
that are available for each individual pulsar. These positions were
then fitted and the results were stored. This procedure was repeated
$10^{5}$ times, yielding distributions as shown in Fig. \ref{fig:bootstrap-histos}.
During the fitting procedure we allow for absolute positional offsets
between the different data sets (typically of the order of several
mas). Such offsets are expected for several reasons: i) the observations
were conducted at different frequencies; ii) the different campaigns
used different calibrator sources, which may have different systematic
errors \citep[e.g.][]{kovalev08,porcas09,sokolovsky11} in their ties
to the International Celestial Reference Frame (ICRF2, \citealt{ma09});
iii) the data were obtained at times that are up to ten years apart
during which improvements to the correlator models and Earth orientation
parameters introduce offsets; and iv) the data were taken with different
instruments (VLBA and EVN) that use different hardware/software correlators.
Table \ref{tab:my-fitting-results} summarizes the estimates of $\mu$
and $\pi$ from the individual fits, from the different combinations
of data sets, and from the bootstrapping method (where we quote the
most compact 68\% confidence interval), as well as the implied pulsar
distances and transverse velocities. The latter are corrected for
solar motion and differential Galactic rotation and refer to the local
standard of rest \foreignlanguage{british}{(LSR)}. Regardless of estimation
strategy and combination of available data, all measured values are
consistent within their uncertainties at the one-sigma level.

For pulsar B1929+10 our results confirm the measurements of \citet{chatterjee04},
especially in combination with the earlier data. For pulsars B2020+28
and B2021+51 our observations indicate that they are located at a
distance of $1.39_{-0.06}^{+0.05}\,$kpc and $1.25_{-0.17}^{+0.14}\,$kpc. Hence, they are about $1.1$ and $0.7\,$kpc
closer to the solar system than what was implied by the measurements
of \citet{brisken02} alone (Table \ref{tab:previous-fitting-results}).
Considering this discrepancy of a factor of about 2, we suspect that
the uncertainties on the position measurements for B2020+28 and B2021+51
as reported by the authors were underestimated. Accordingly, we did
not include these data in the further analysis. Thus, in the following,
for B1929+10 we adopted the astrometric parameters obtained from bootstrapping
all available data, while for B2020+28 and B2021+51 we used the parameters
as measured with our new 5$\,$GHz data alone. Hence,  the analysis
below is based on the astrometry as listed in Table \ref{tab:sim-parameters}.
\begin{figure*}
\begin{centering}
\includegraphics[clip,width=0.9\textwidth]{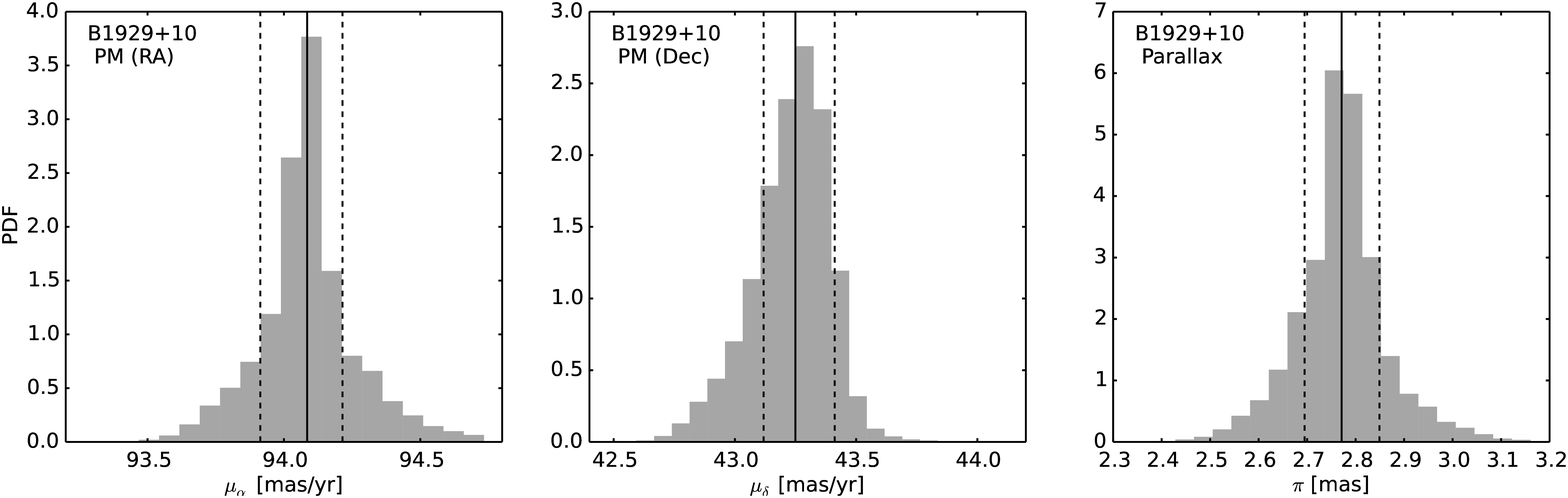}
\par\end{centering}

\begin{centering}
\includegraphics[clip,width=0.9\textwidth]{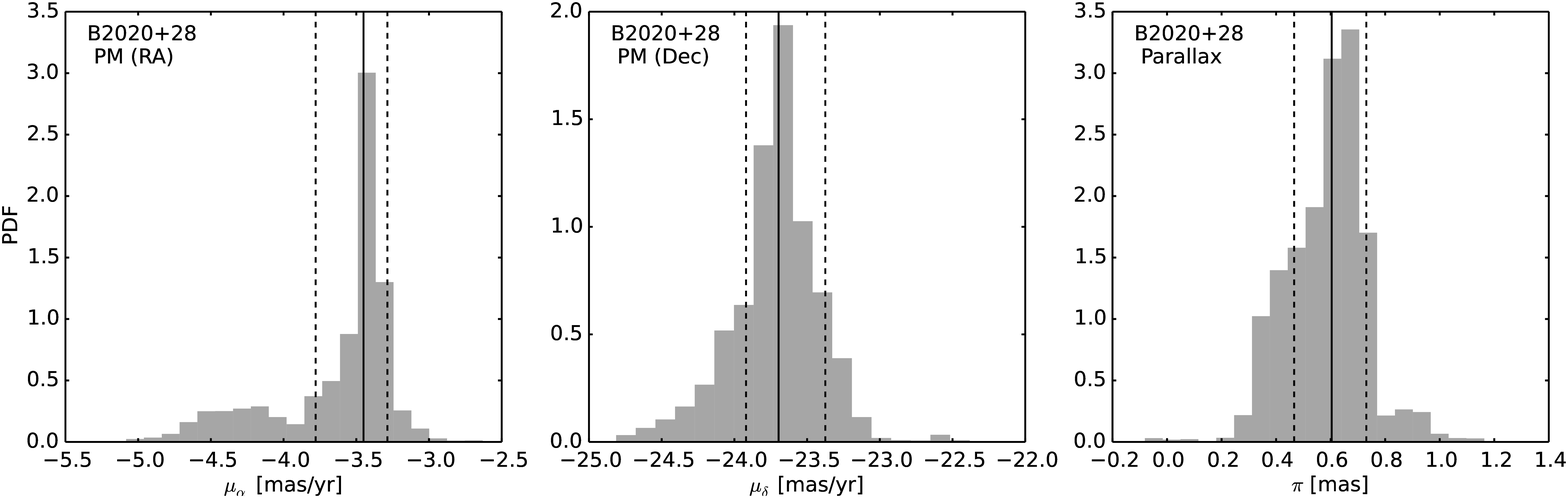}
\par\end{centering}

\begin{centering}
\includegraphics[clip,width=0.9\textwidth]{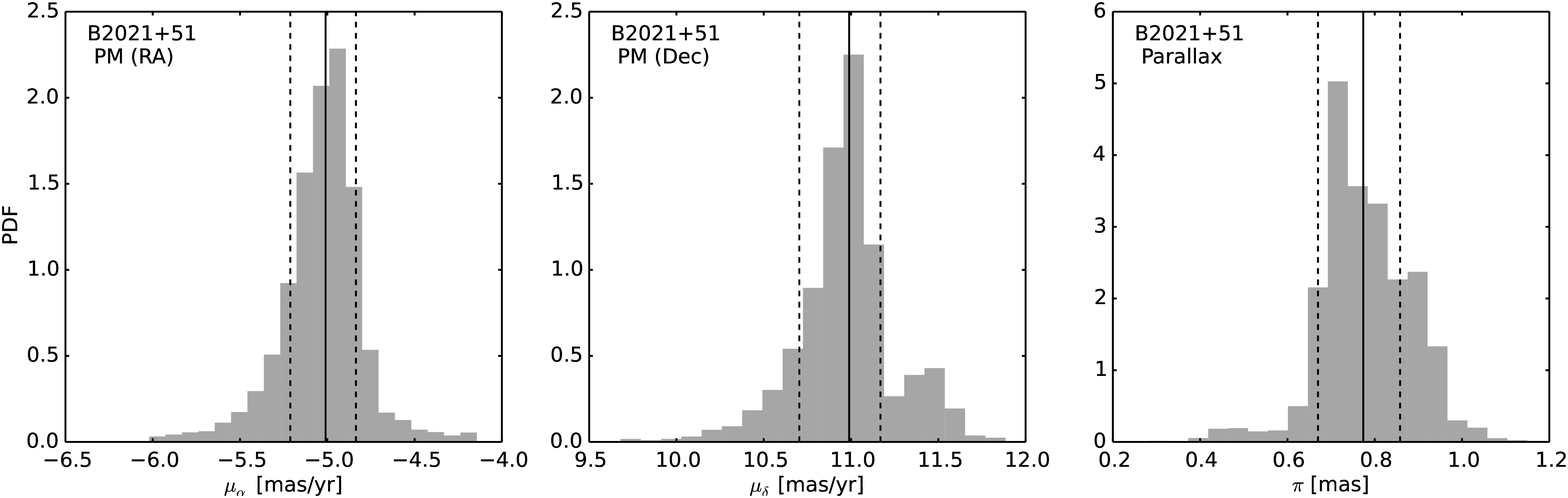}
\par\end{centering}

\caption[Probability density functions of fitting results from bootstrapping
all available data]{\label{fig:bootstrap-histos}Probability density functions of fitting
results from bootstrapping all available data. From top to bottom:
B1929+10, B2020+28, and B2021+51. Columns are from left to right: proper
motion in RA, proper motion in Dec, and parallax. The solid and dashed
vertical lines indicate the median and the most compact 68\% confidence
intervals as listed in Table \ref{tab:my-fitting-results}.}
\end{figure*}
\begin{figure*}
\begin{centering}
\subfloat{\includegraphics[width=0.25\textwidth]{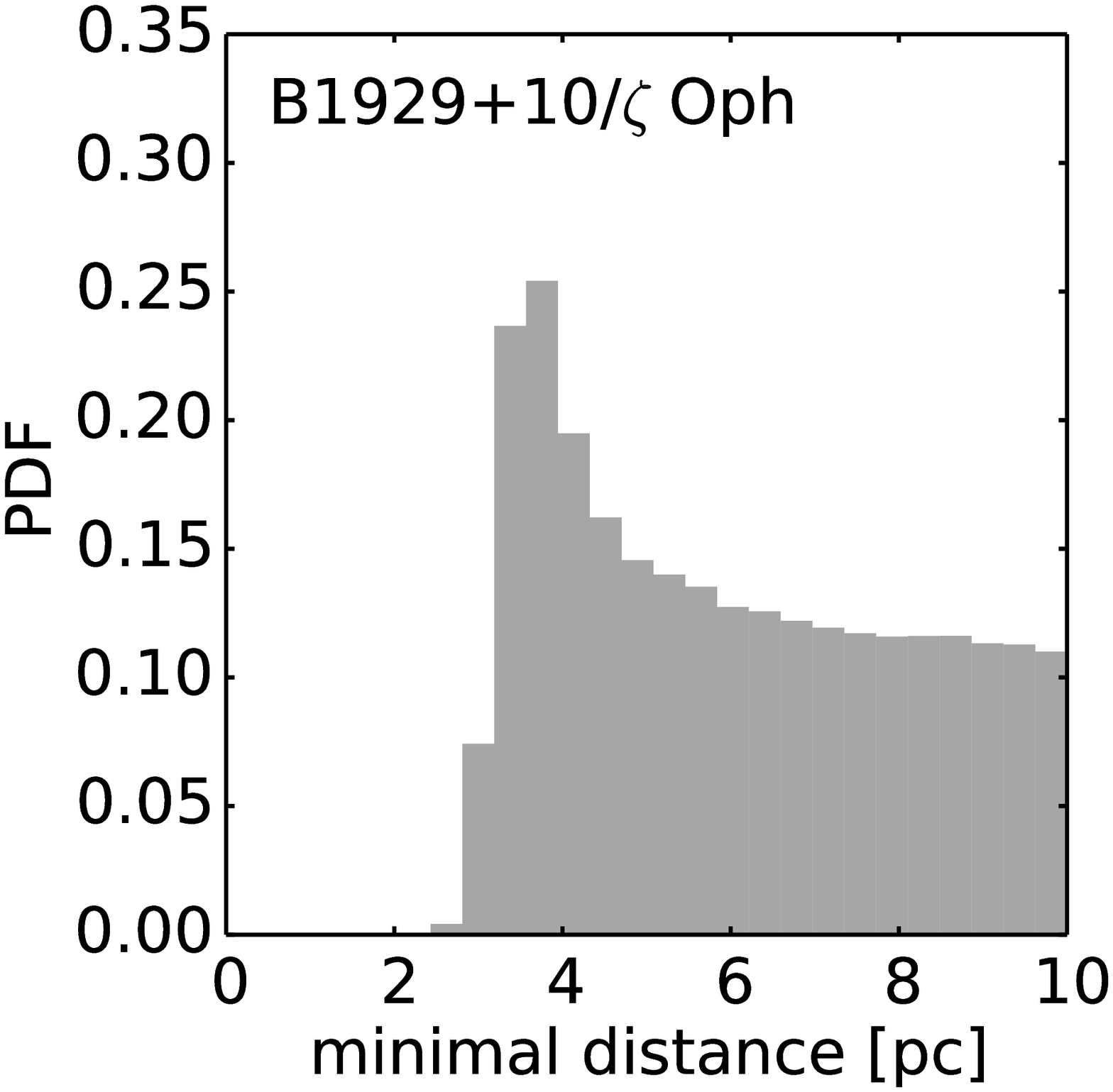}}\subfloat{\includegraphics[width=0.25\textwidth]{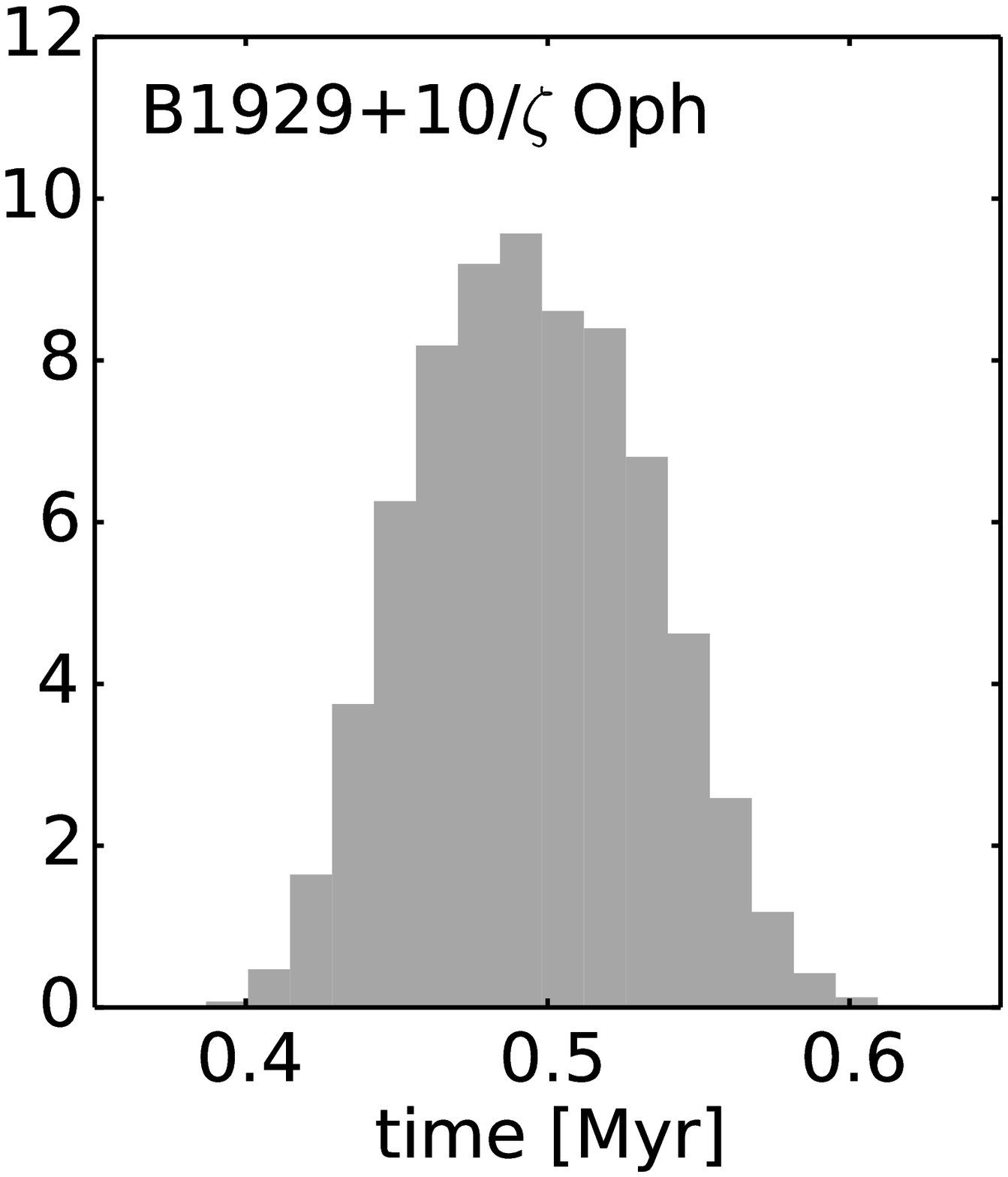}}\subfloat{\includegraphics[width=0.25\textwidth]{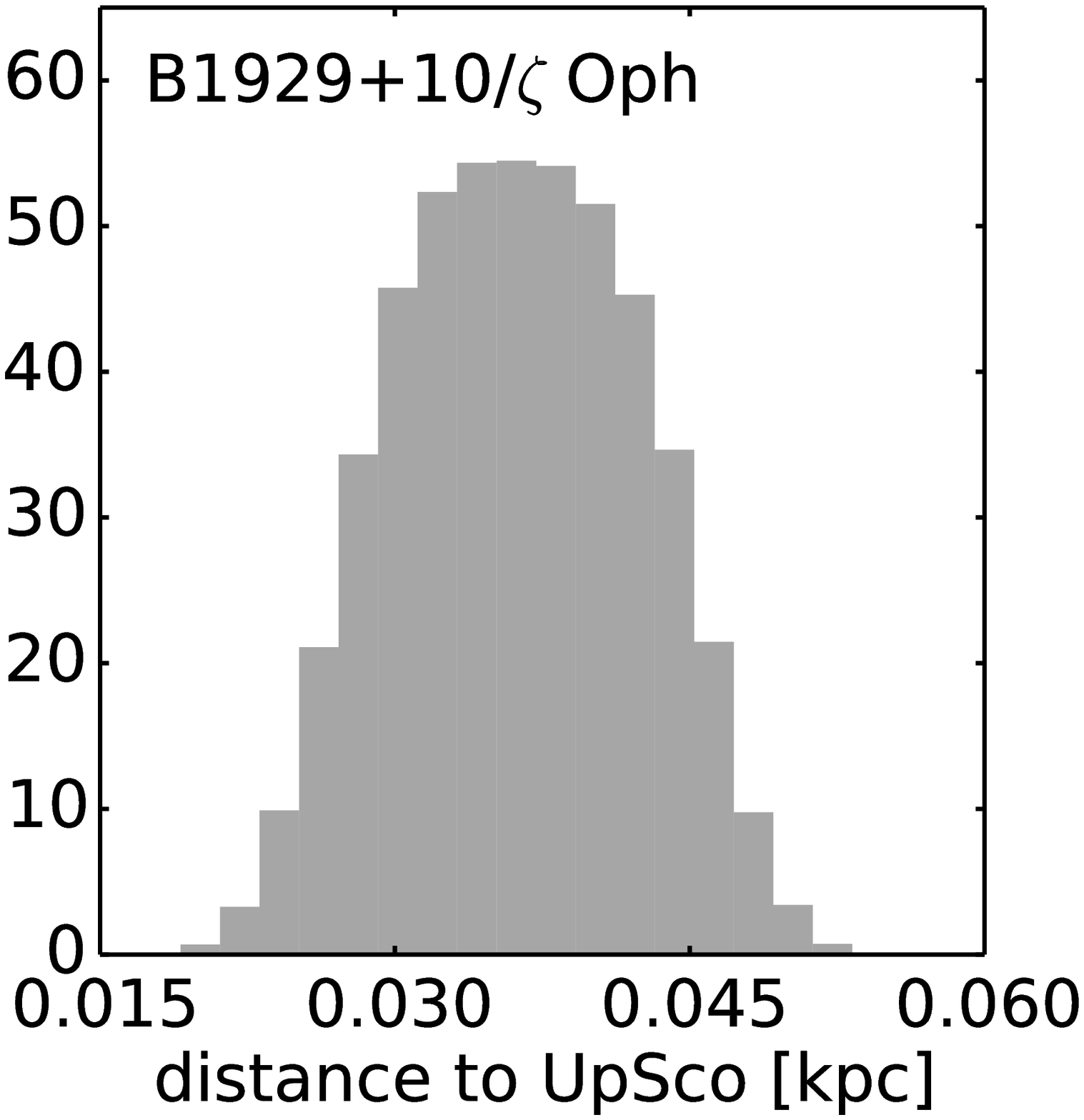}}
\par\end{centering}

\begin{centering}
\vspace{-0.5cm}\subfloat{\includegraphics[width=0.25\textwidth]{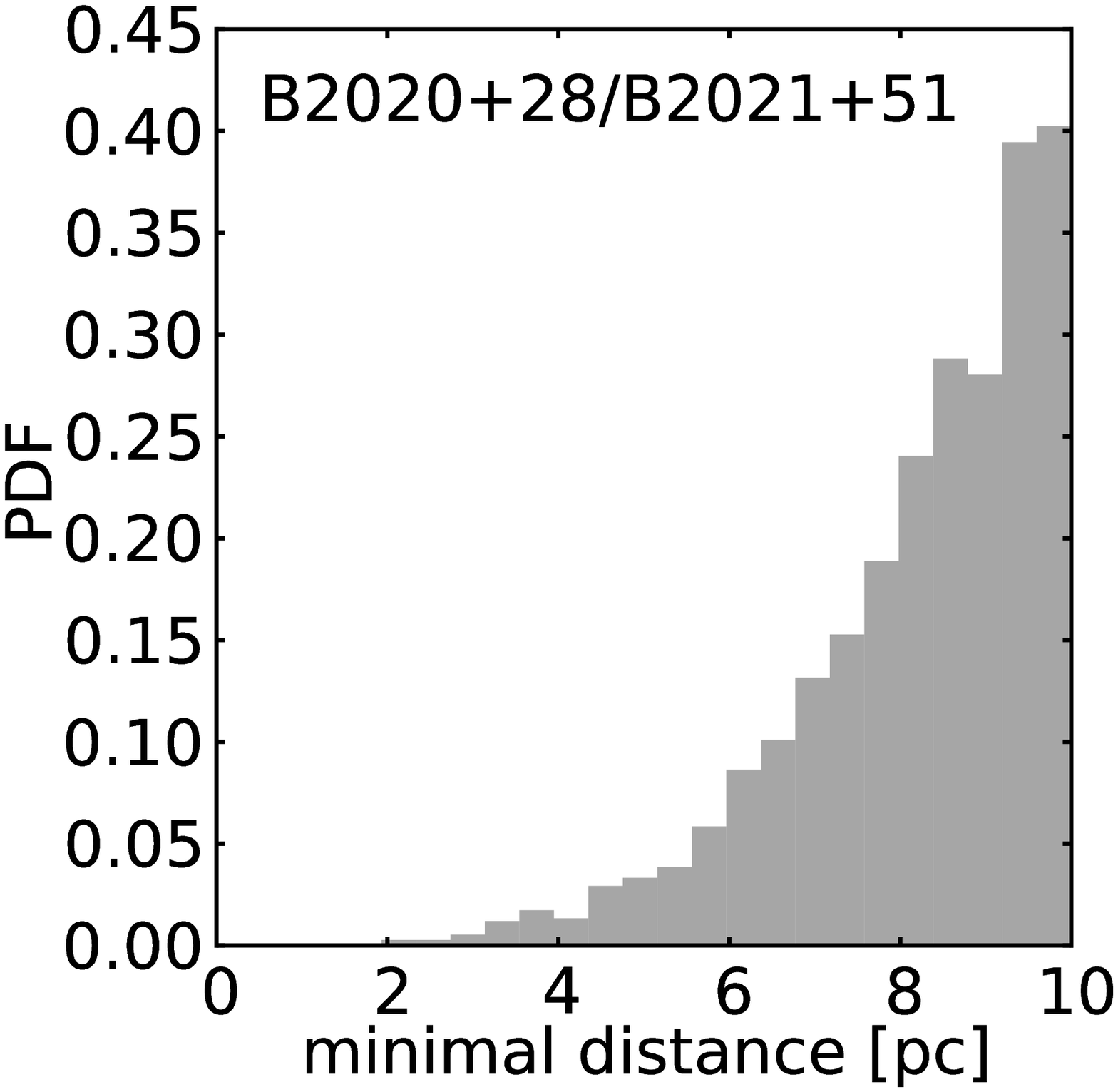}}\subfloat{\includegraphics[width=0.25\textwidth]{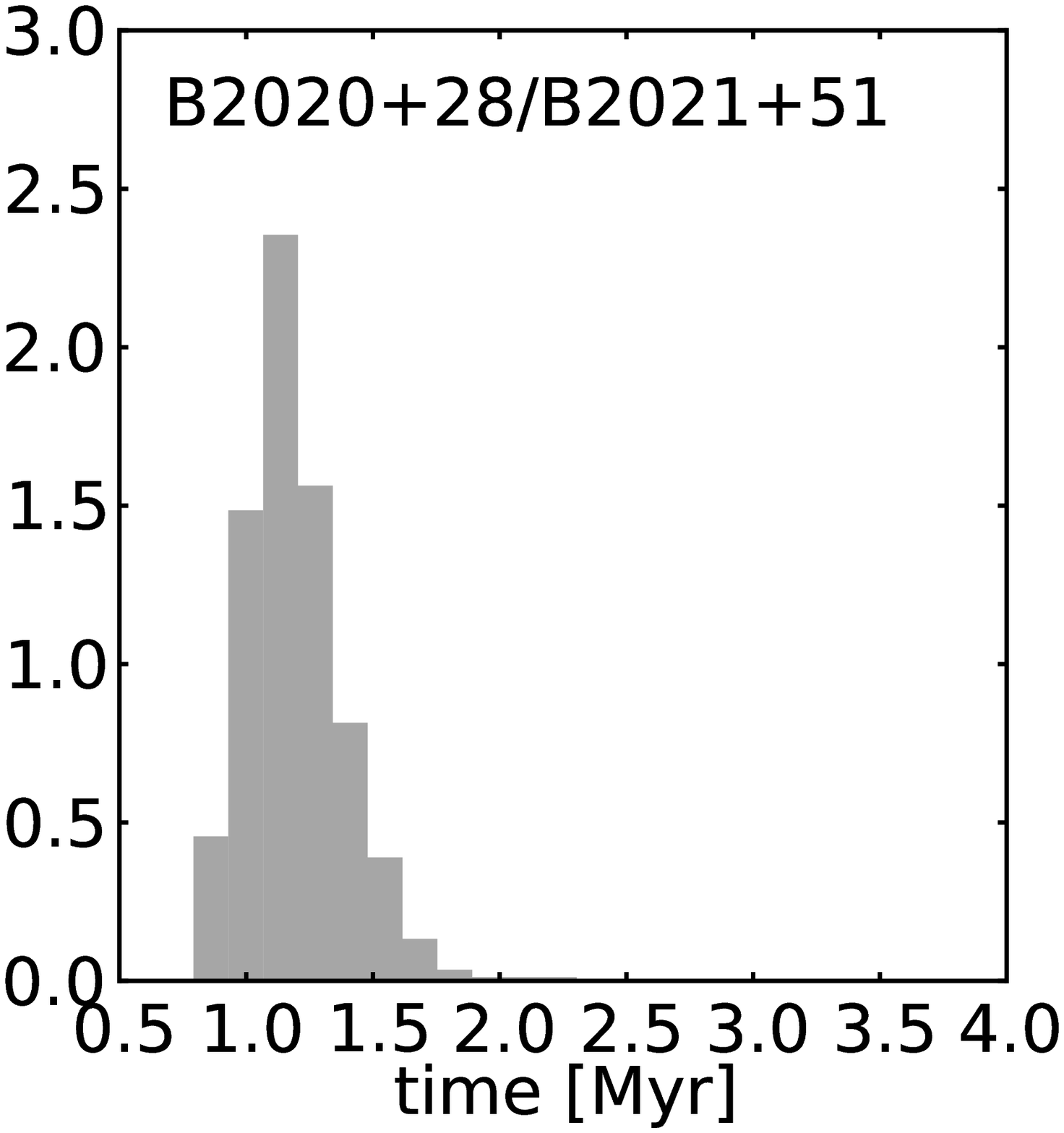}}\subfloat{\includegraphics[width=0.25\textwidth]{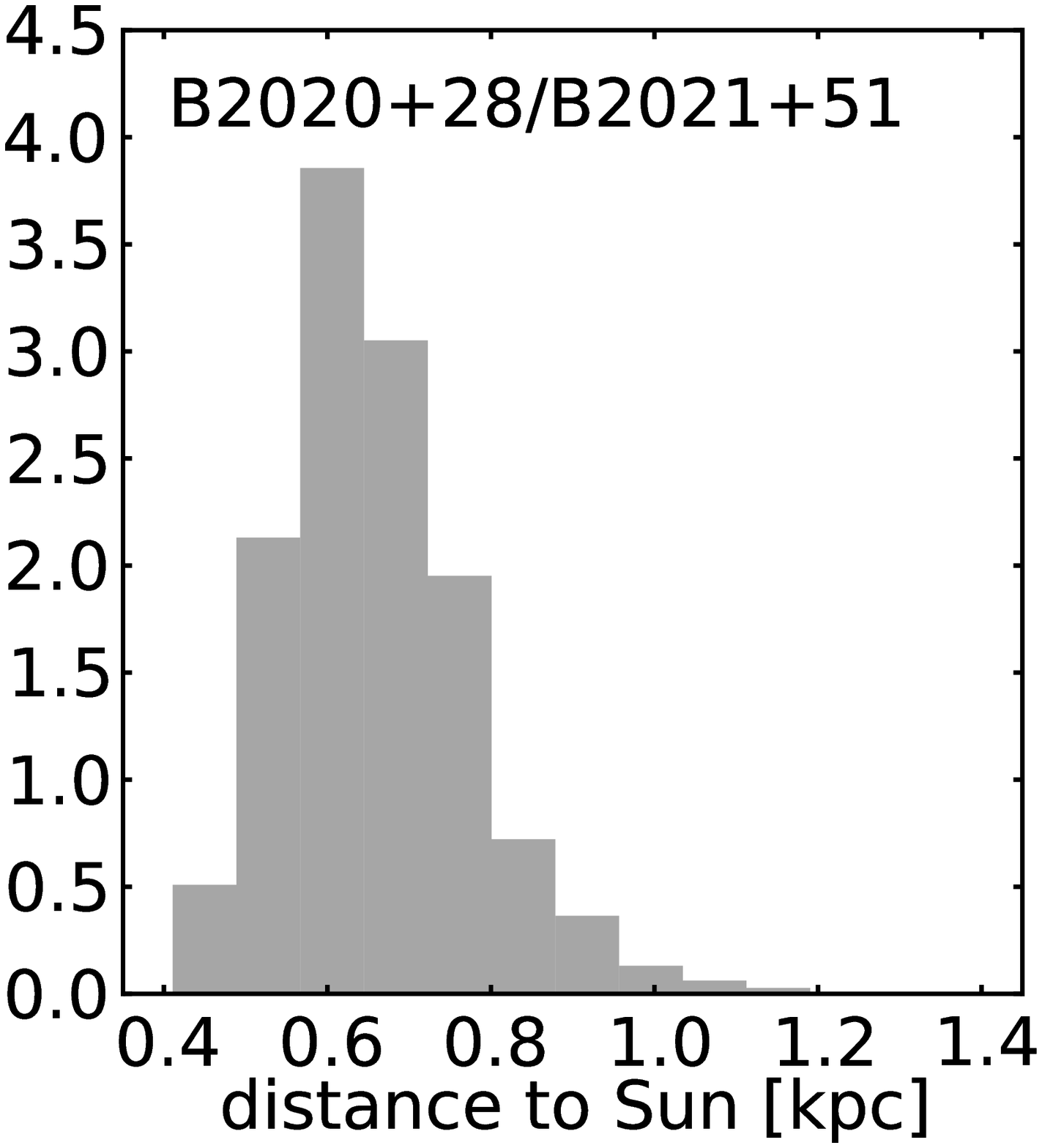}}
\par\end{centering}

\caption[Probability density functions of minimal distances, time of minimal
approach and distance to Upper Scorpius and the Sun for B1929+10/$\zeta$
Oph, and B2020+28/B2021+51]{\label{fig:distance-time-histos}Probability density functions of
minimal distances (left column), time of minimal approach (middle
column), and distance to Upper Scorpius and the Sun for B1929+10/$\zeta$
Oph (upper row) and B2020+28/B2021+51 (lower row), respectively. The
stellar pair to which the figures apply is indicated in the top of
each panel.}
\end{figure*}

\section{Simulations of pulsar orbits}

\begin{table}[!b]
\caption[\selectlanguage{british}%
Astrometry used in simulations\selectlanguage{english}%
]{\label{tab:sim-parameters}\foreignlanguage{british}{}Astrometric
parameters used in the simulations }

\begin{centering}
\begin{tabular}{lccc}
\noalign{\vskip-0.3cm}
\hline\hline &  &  & \tabularnewline
\noalign{\vskip-0.2cm}
 & $\mu_{\alpha}$ & $\mu_{\delta}$ & $\pi$\tabularnewline
Source & {[}mas$\,$yr$^{-1}${]} & {[}mas$\,$yr$^{-1}${]} & {[}mas{]}\tabularnewline
\hline 
\noalign{\vskip\doublerulesep}
B1929+10 & $94.08\pm0.17$ & $43.25\pm0.16$ & $2.77\pm0.08$\tabularnewline
\noalign{\vskip\doublerulesep}
$\zeta$ Oph\tablefootmark{a} & $15.26\pm0.26$ & $24.79\pm0.22$ & $8.91\pm0.20$\tabularnewline
\noalign{\vskip\doublerulesep}
B1952+29\tablefootmark{b} & $-30\pm6$ & $-34\pm8$ & $1.4\pm1.0$\tabularnewline
B2020+28 & $-3.34\pm0.05$ & $-23.65\pm0.11$ & $0.72\pm0.03$\tabularnewline
B2021+51 & $-5.08\pm0.42$ & $10.84\pm0.25$ & $0.80\pm0.11$\tabularnewline
\hline\vspace{-0.4cm} &  &  & \tabularnewline
\end{tabular}
\par\end{centering}

\tablefoot{
\tablefoottext{a}{From van Leeuwen (2007)}
\tablefoottext{b}{Proper motion from Hobbs et al. (2004), parallax from the ATNF Pulsar Catalogue (Manchester et al., 2005).}
}
\end{table}
To shed new light on possible common origins of B1929+10/$\zeta$ Oph and B2020+28/B2021+51, we used the pulsar astrometric
parameters described above and traced their orbits back in time through
the Galactic potential. For the runaway star $\zeta$ Oph we used the
latest proper motion and parallax measurements from \citet{vanLeeuwen07}
(Table \ref{tab:sim-parameters}) and adopted the value for the radial
velocity $V_{\text{rad}}=-9.0\pm5.5\,$km$\,$s$^{-1}$ from \citet{kharchenko07}.
Our astrometric measurements yield information about the transverse
motion of the pulsars, but they do not contain any information about
the radial velocity. To estimate the full 3D velocity
vector, we simulated the possible radial component from our measured
transverse components and the space velocity distribution of young
pulsars as empirically derived by \citet{hobbs05}. To account for
the uncertainties of the measured parameters $\mu_{\alpha},\,\mu_{\delta},\,\pi$,
and the unknown radial velocity, we assumed that all parameters
are distributed normally (where the half-width of the Gaussian is
given by the higher absolute value of the upper and lower errors of
the bootstrapping results) and performed three million Monte Carlo simulations.
The obtained velocity vectors were corrected for the solar motion with
respect to the LSR, for differential Galactic rotation, and also for
the velocity of the LSR. The Galactic potential we used in our simulations
is the potential that was described in full detail in \citet{vlemmings04}\foreignlanguage{british}{,}
the main parameters of which we summarize here in brief. For
consistency reasons, the pulsar orbits were traced back through the
same Stäckel potential as in \citet{vlemmings04}; consisting of
a thin disk, a thick disk, and a halo component whose axis ratios
are $75.0,$ $1.5,$ and $1.02$, respectively. We kept the relative
contributions of thin and thick disks and of the halo at $0.13,$
$0.01,$ and $1.0$, respectively. For a complete description of each
parameter of the Stäckel potential, we refer to \citet{famaey03}.
We adopted parameters for solar motion from \citet{schoenrich12}: $R_{\odot}=8.27\,\text{kpc and }(U,\, V,\, W)=(13.84,\,12.24,\,6.1)\,\text{km}\,\text{s}^{-1}$.
Each object's trajectory was sampled at time intervals of $10^{3}\,$yr
using a fourth-order Runge-Kutta numerical integration method. For
each time step the distances between the two objects under consideration
were computed within the Galactic reference frame, and we recorded only the simulation
input parameters of trajectories that resulted in a minimum distance
of less than 10$\,$pc. In addition to the separation
between the individual objects, we also computed their distances to
the Sun (B2020+28/B2021+51) and to the Upper Scorpius region (B1929+10/$\zeta$ Oph). To compute the latter, we traced the trajectory of Upper
Scorpius back in time using the astrometric values as listed in Table
2 of \citet{deZeeuw99}.\\
\\
For consistency checks we used the input parameters of \citet{hoogerwerf01}
to compute the trajectories of B1929+10 and $\zeta$ Oph. In total,
37521 of the three million sampled trajectories ($1.2$\%) cross within
$10\,$pc of each other. This is close to the percentage found in
\citet{hoogerwerf01}: 30822 out of three million, or 1.0\%. The smallest
separation we found is $0.19\,$pc (compared to $0.35\,$pc). Furthermore,
while \citet{hoogerwerf01} reported that in 4214 (0.14\%) simulations
the trajectories of both the pulsar and the runaway star not only
pass within $10\,$pc of each other, but also pass within less than
$10\,$pc of Upper Scorpius, we found that 6816 (0.23\%) of our simulations
meet these conditions. The differences in the results are probably
due to the different set-ups of the Galactic potentials. We did
reproduce the general trend found by \citet{hoogerwerf01}, however. When we
ran the simulations using the same input parameters for B1929+10 as
\citet{hoogerwerf01}, but used the latest parameters for $\zeta$ Oph
from \citet{vanLeeuwen07}, a total of 82840 (2.7\%) simulated orbits
cross within $10\,$pc, in only 8 of which both the pulsar and the
star are less than $10\,$pc away from Upper Scorpius.

\begin{figure*}
\begin{centering}
\subfloat{\includegraphics[width=0.25\textwidth]{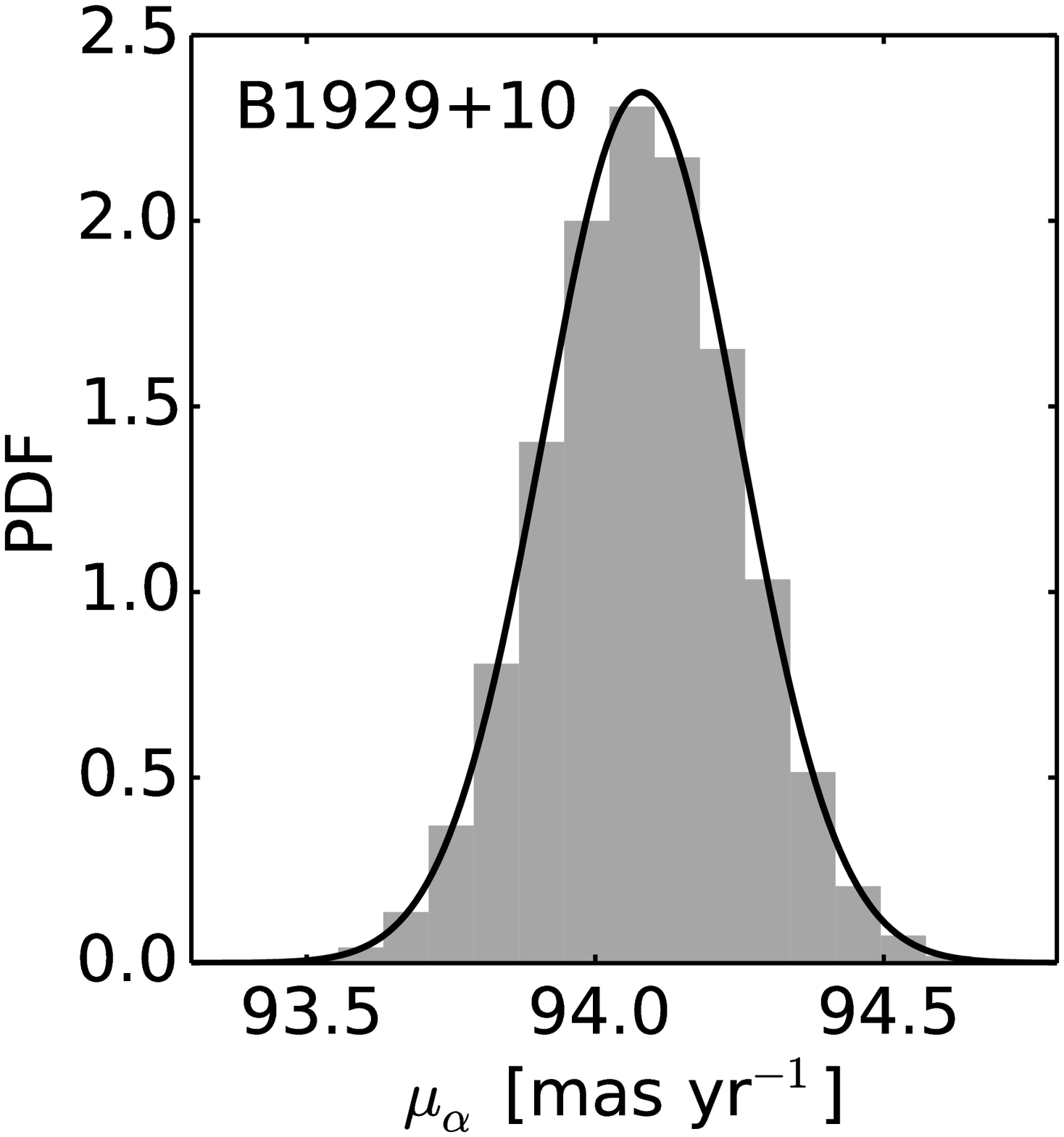}}\subfloat{\includegraphics[width=0.25\textwidth]{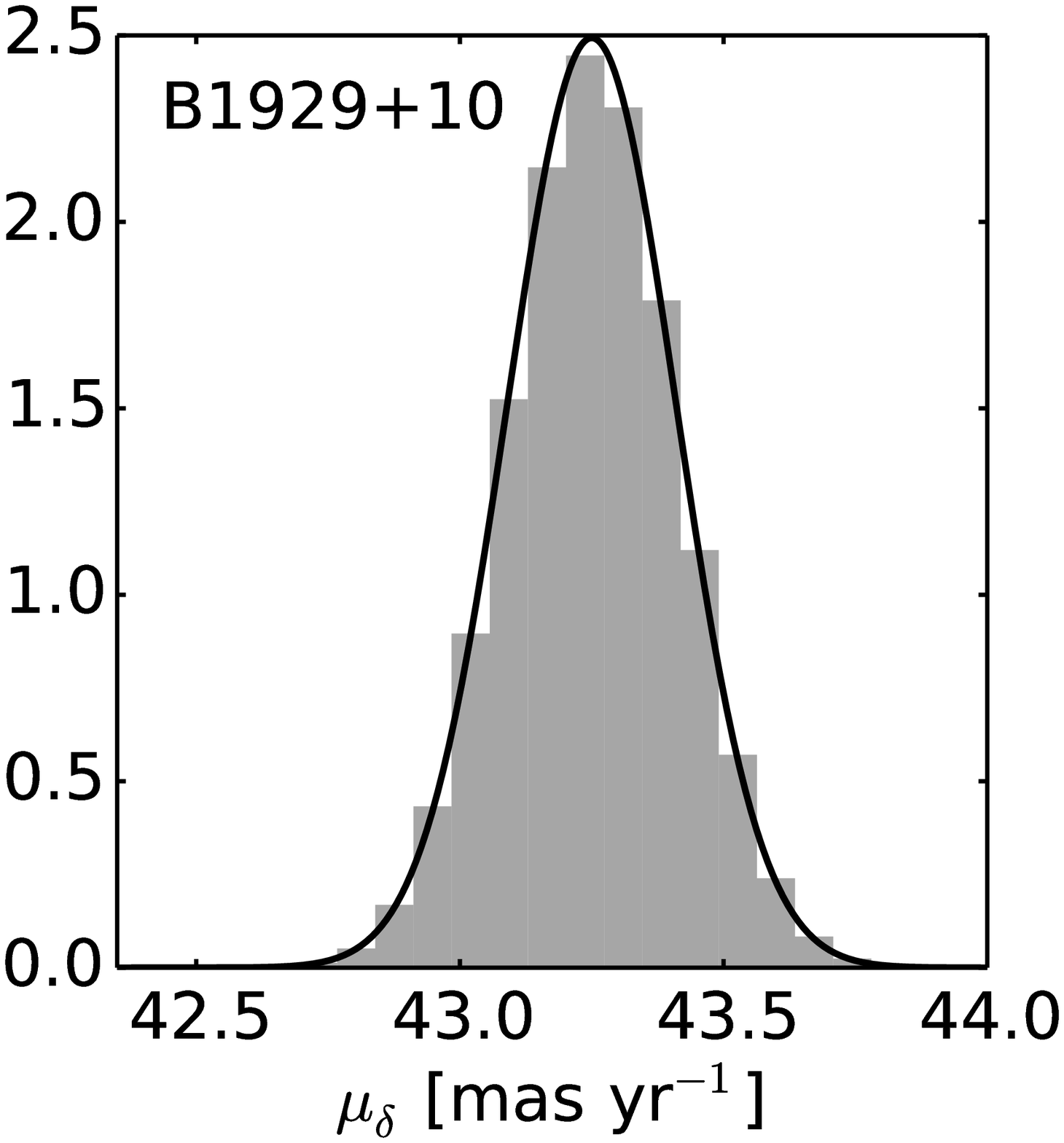}}\subfloat{\includegraphics[width=0.25\textwidth]{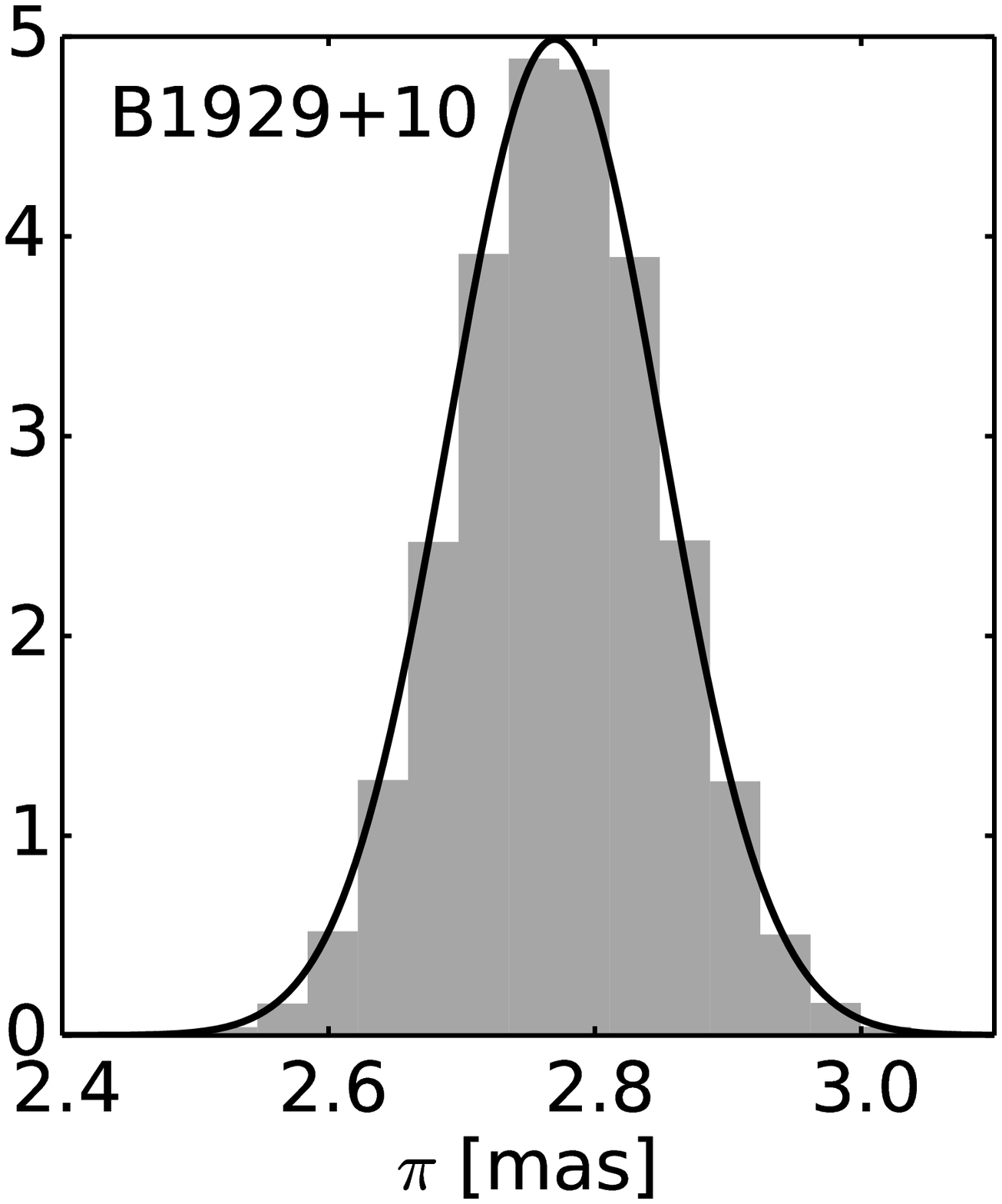}}\subfloat{\includegraphics[width=0.25\textwidth]{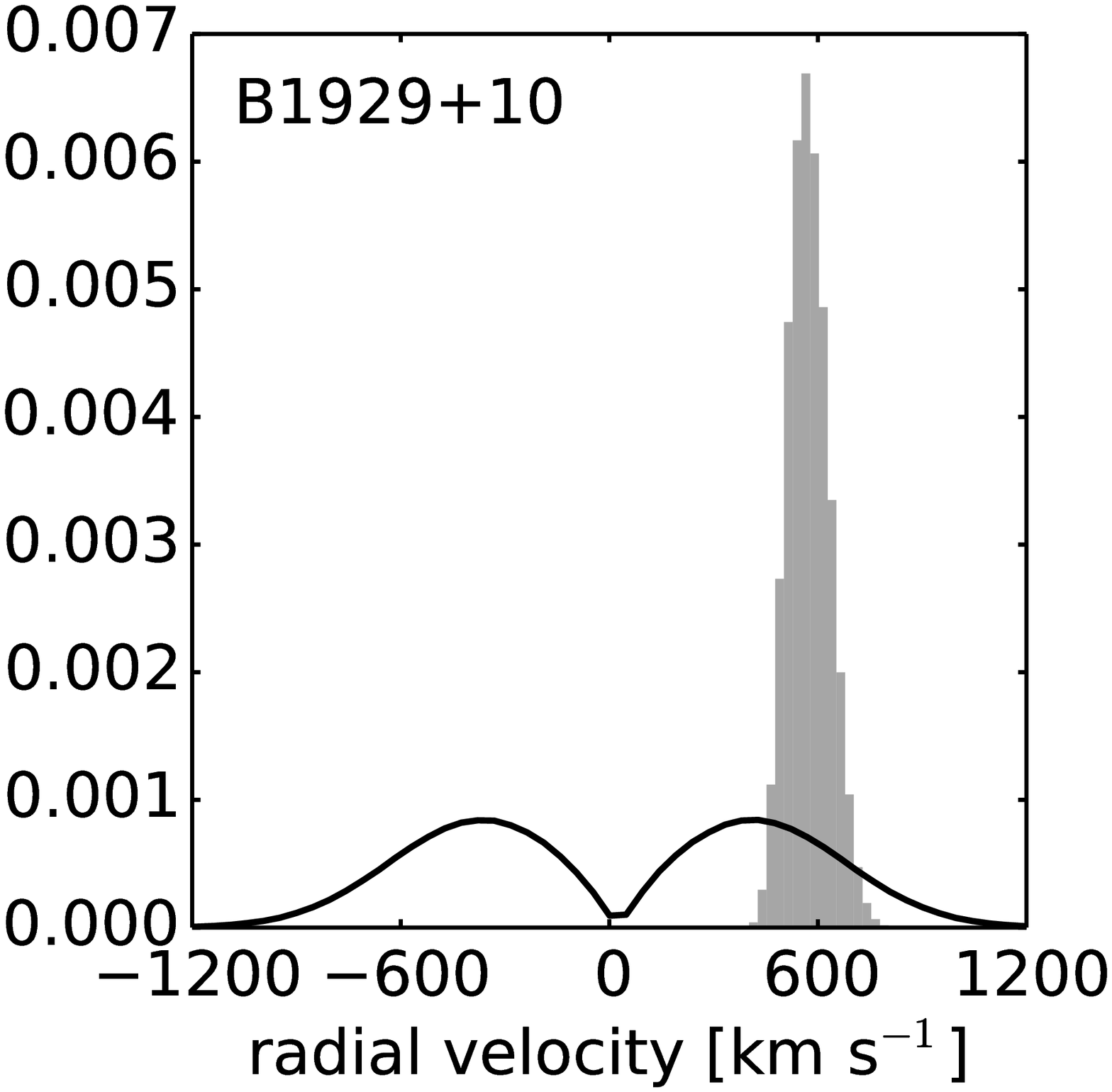}}
\par\end{centering}

\begin{centering}
\vspace{-0.5cm}\subfloat{\includegraphics[width=0.25\textwidth]{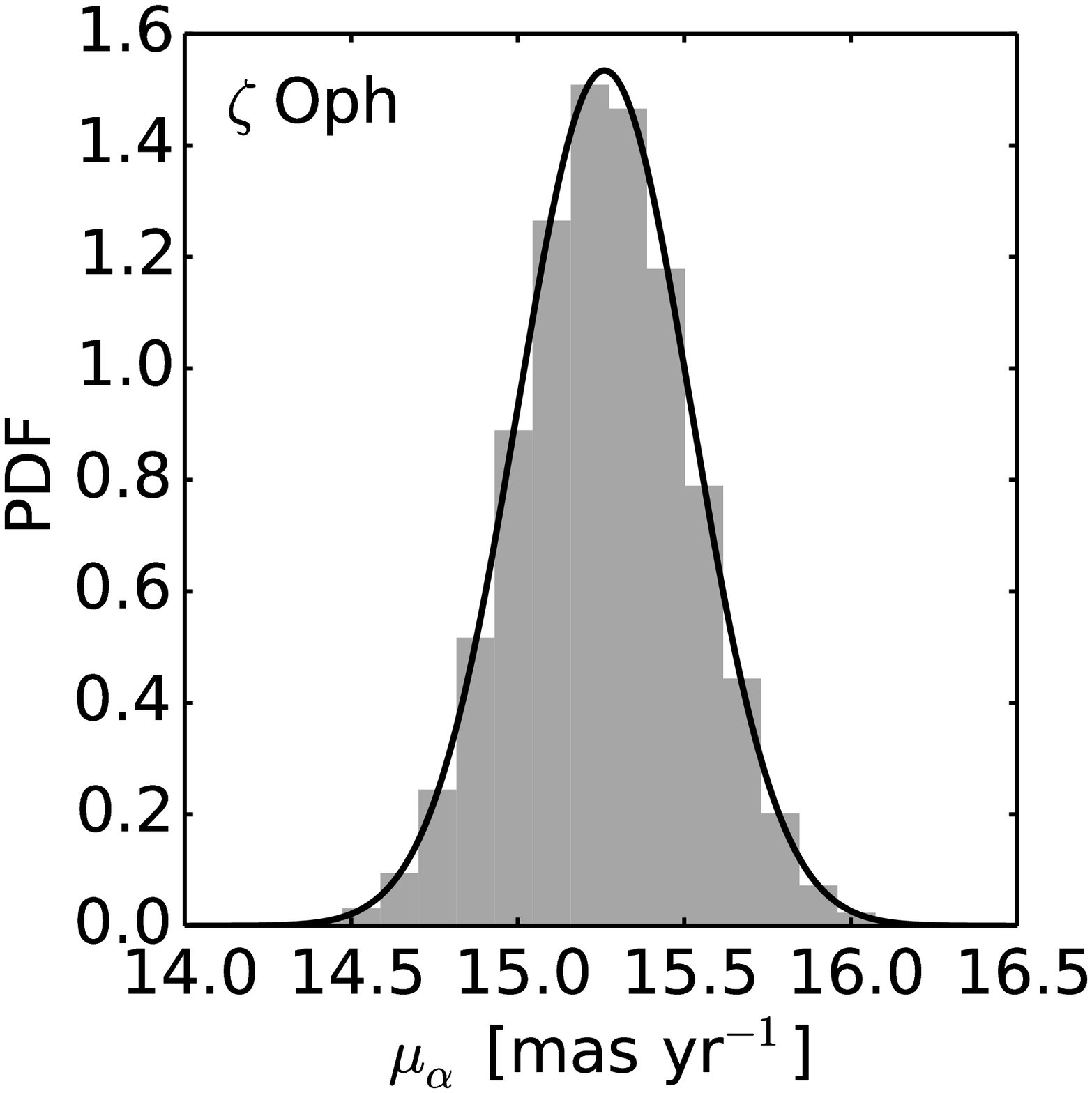}}\subfloat{\includegraphics[width=0.25\textwidth]{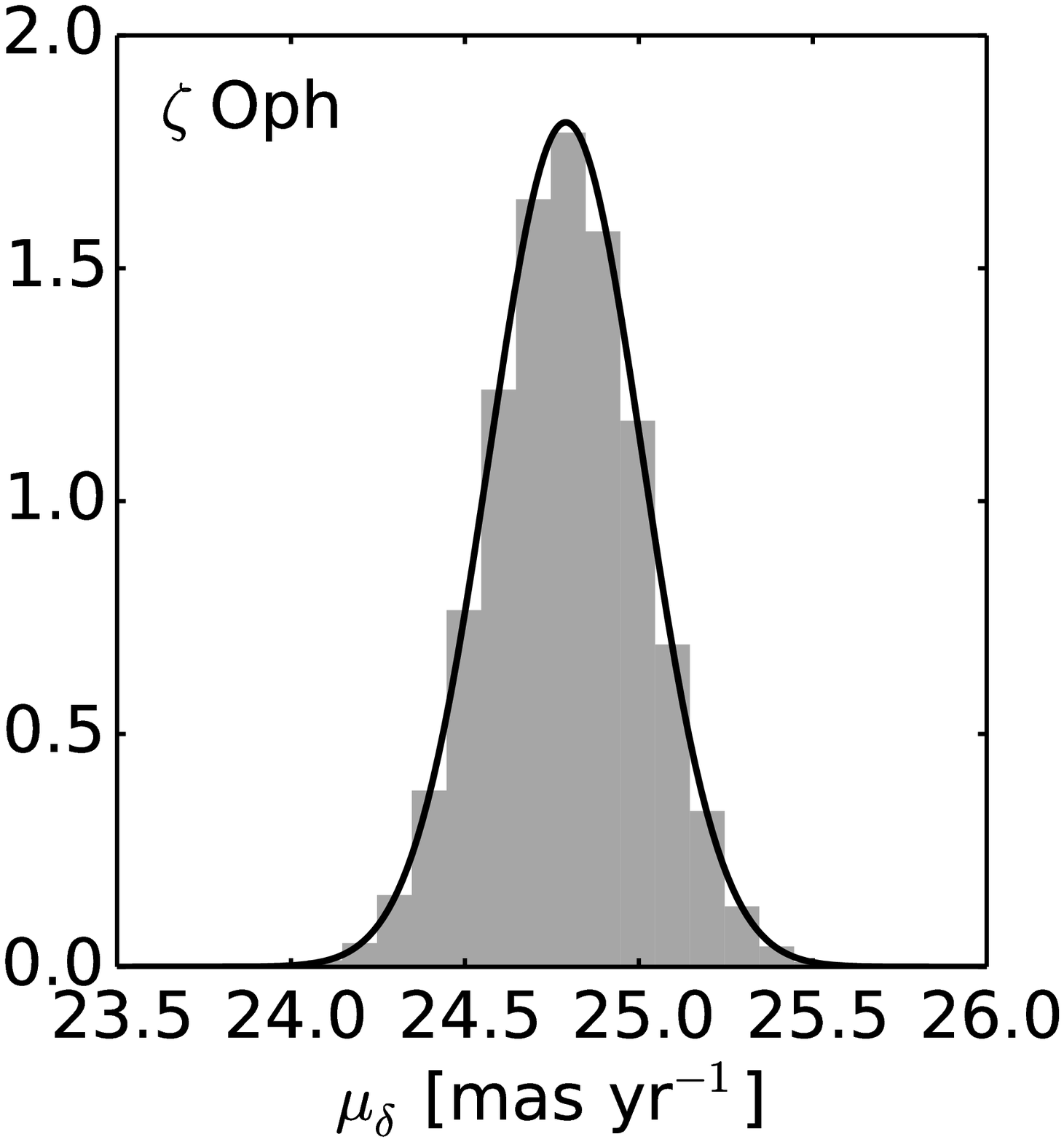}}\subfloat{\includegraphics[width=0.25\textwidth]{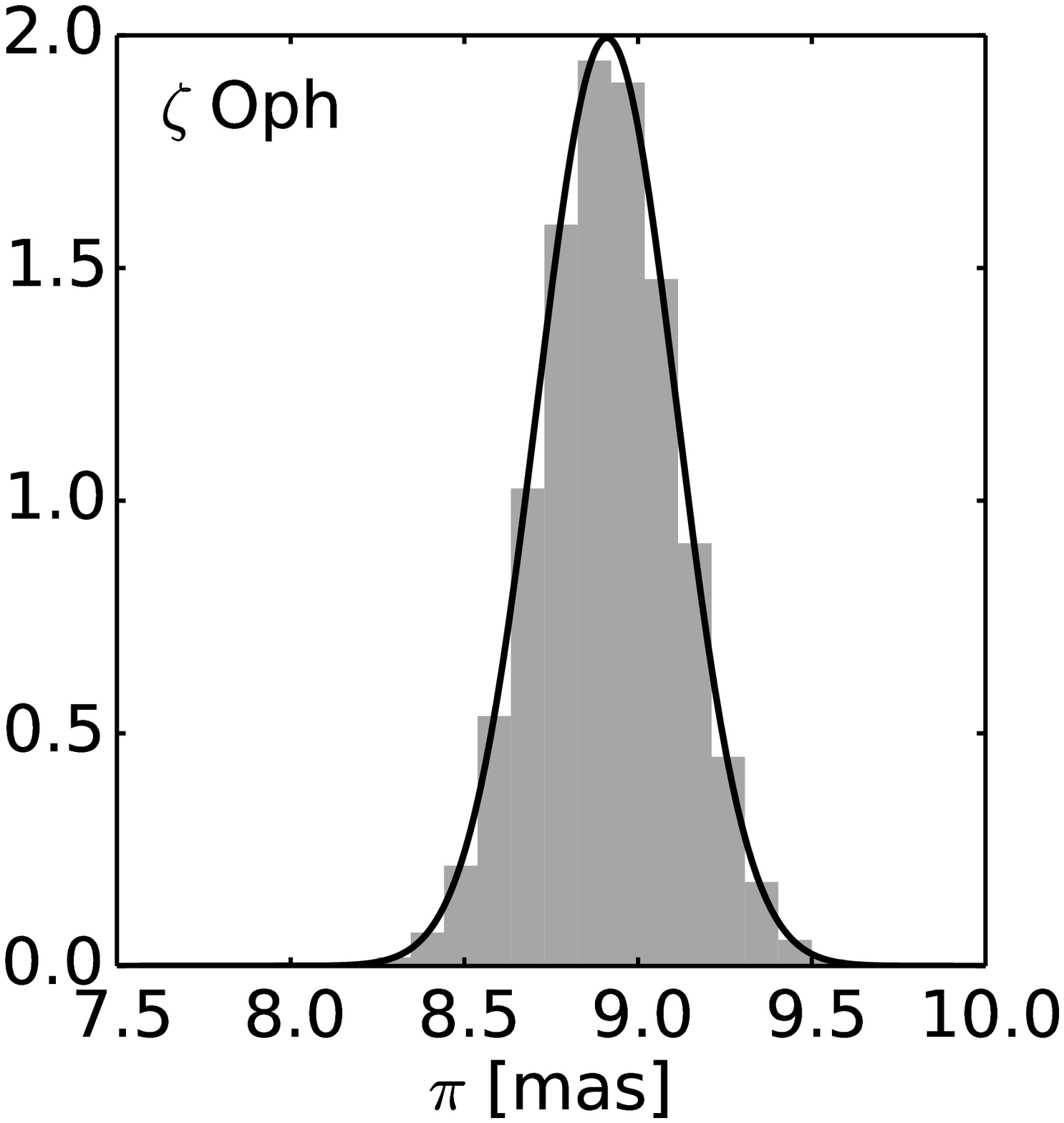}}\subfloat{\includegraphics[width=0.25\textwidth]{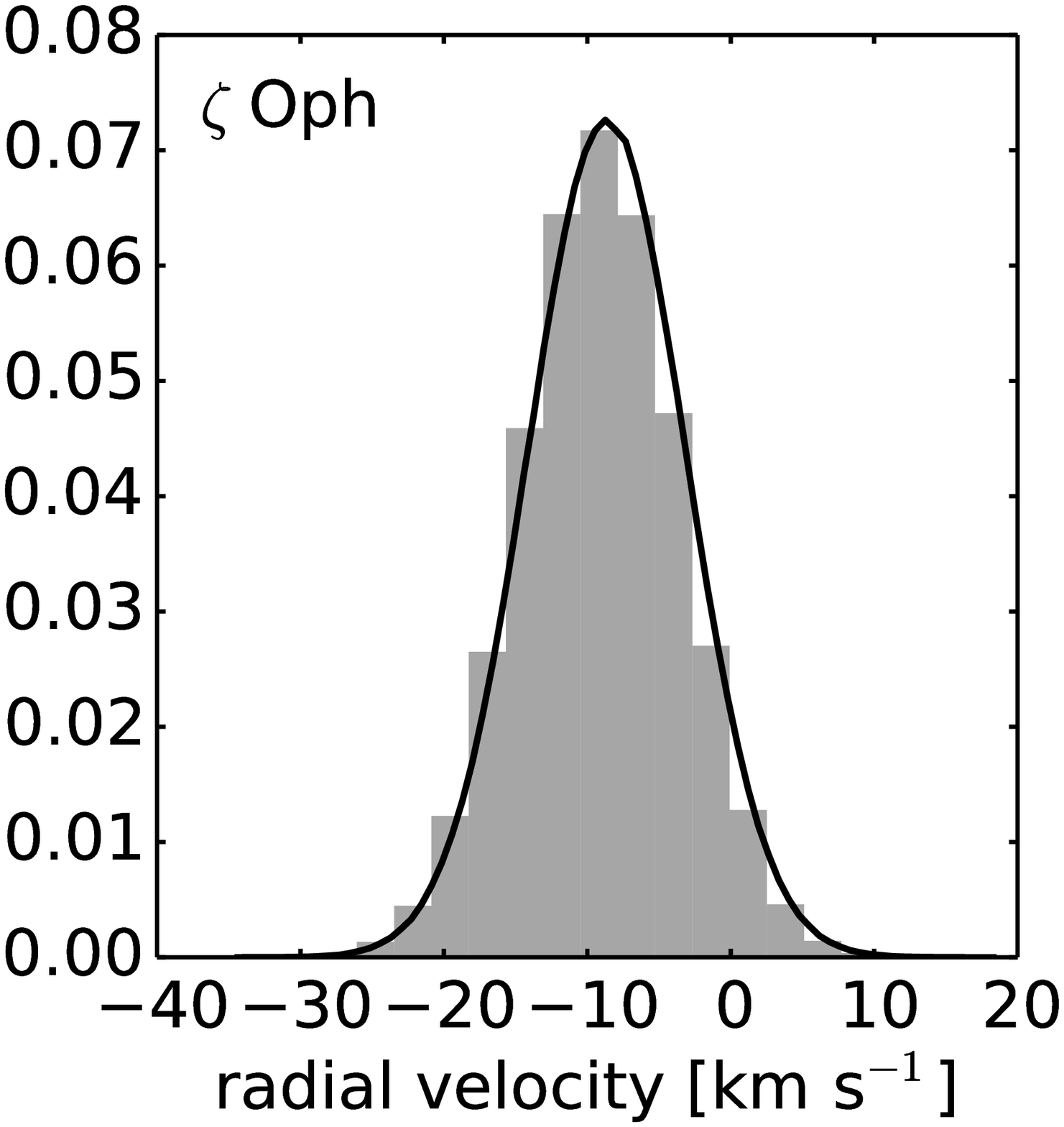}}
\par\end{centering}

\begin{centering}
\vspace{-0.5cm}\subfloat{\includegraphics[width=0.25\textwidth]{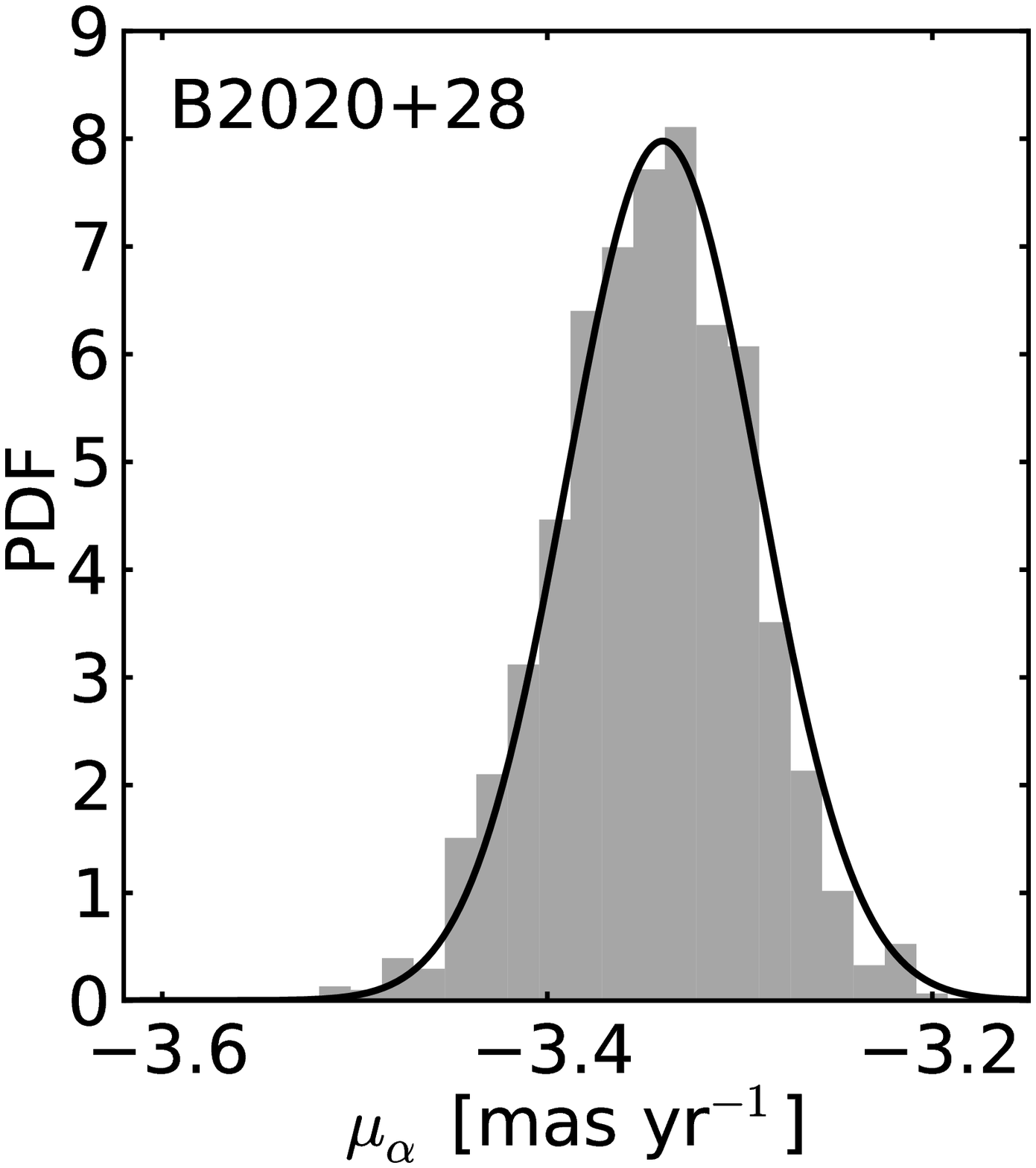}}\subfloat{\includegraphics[width=0.25\textwidth]{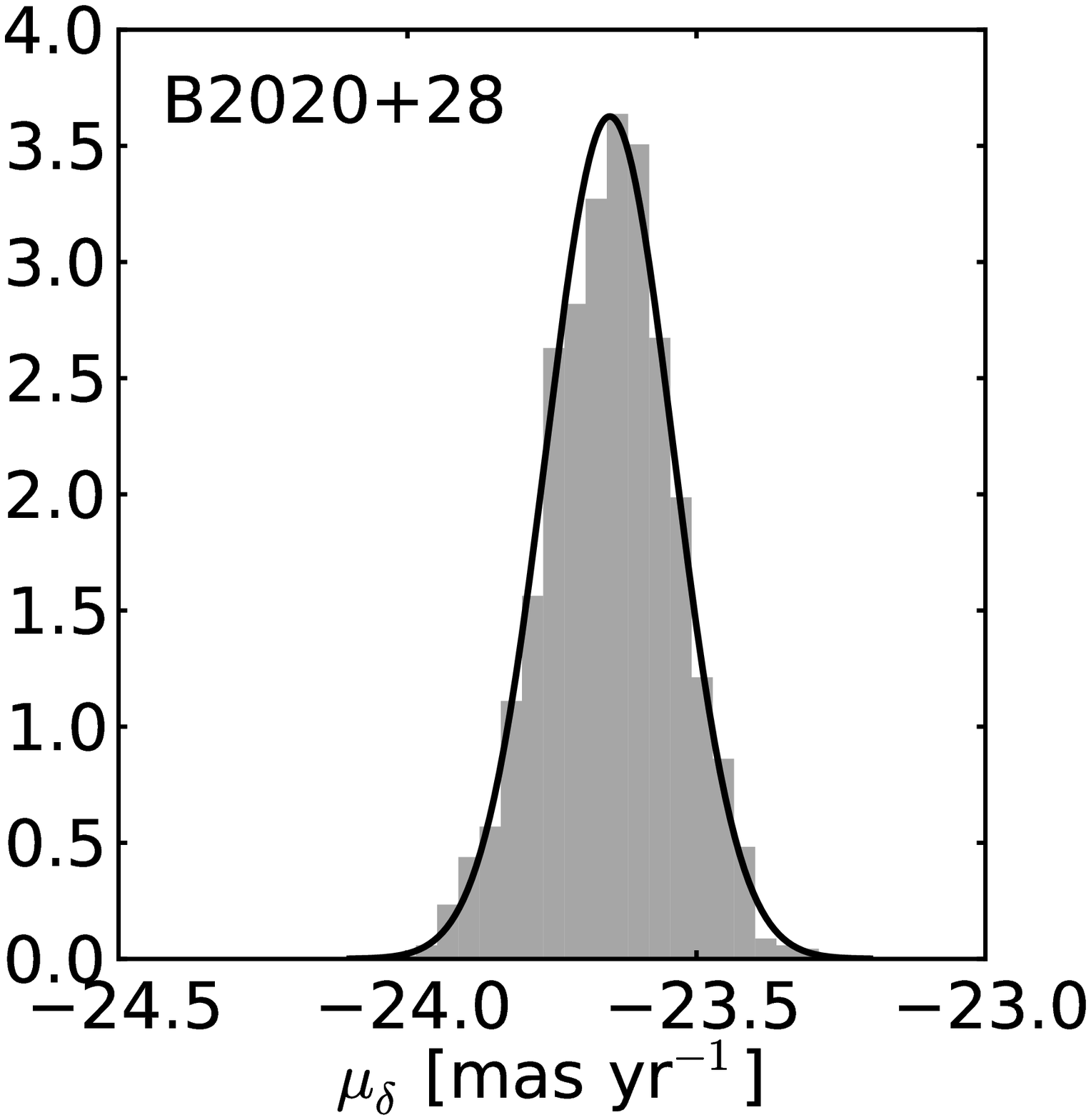}}\subfloat{\includegraphics[width=0.25\textwidth]{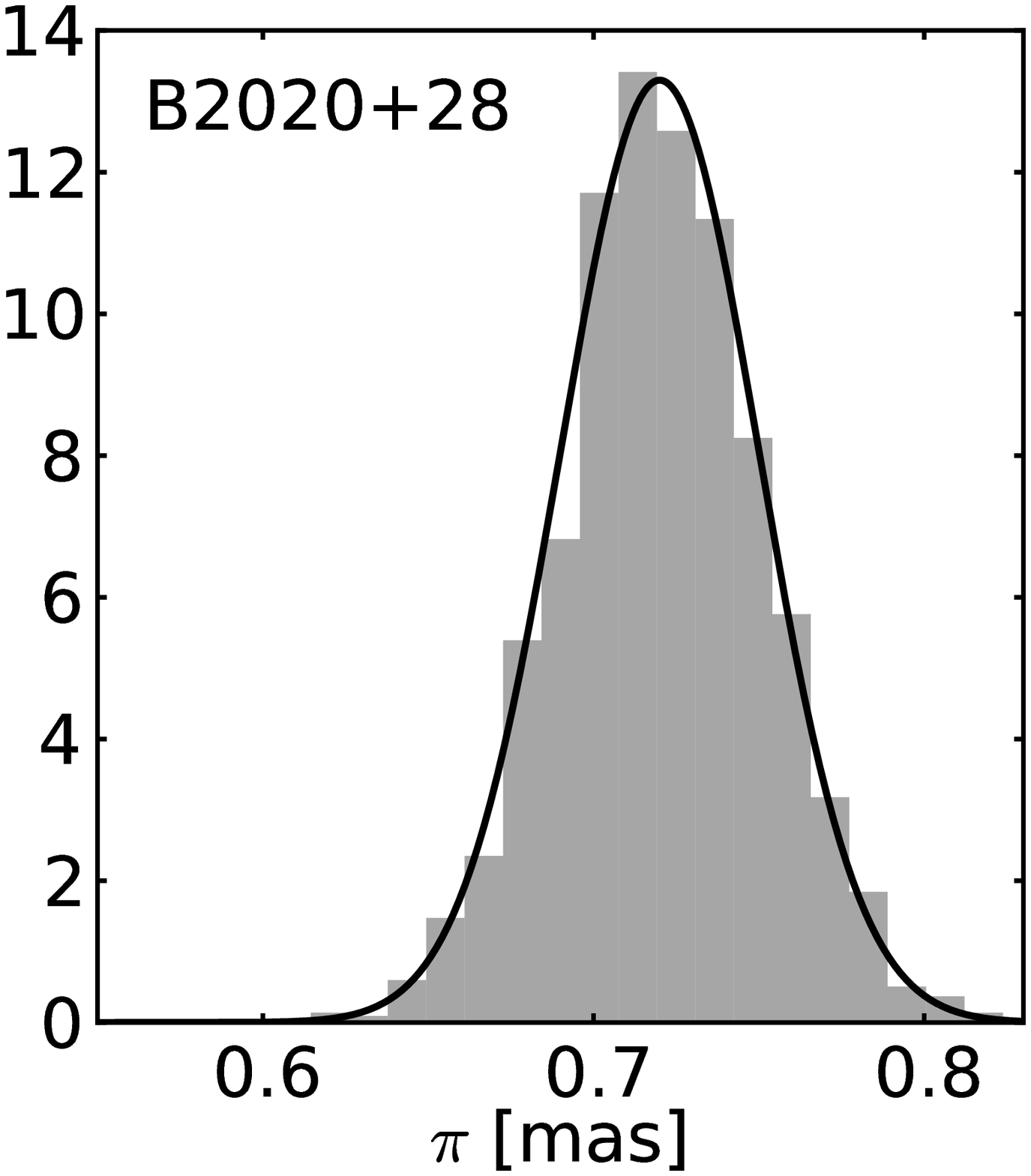}}\subfloat{\includegraphics[width=0.25\textwidth]{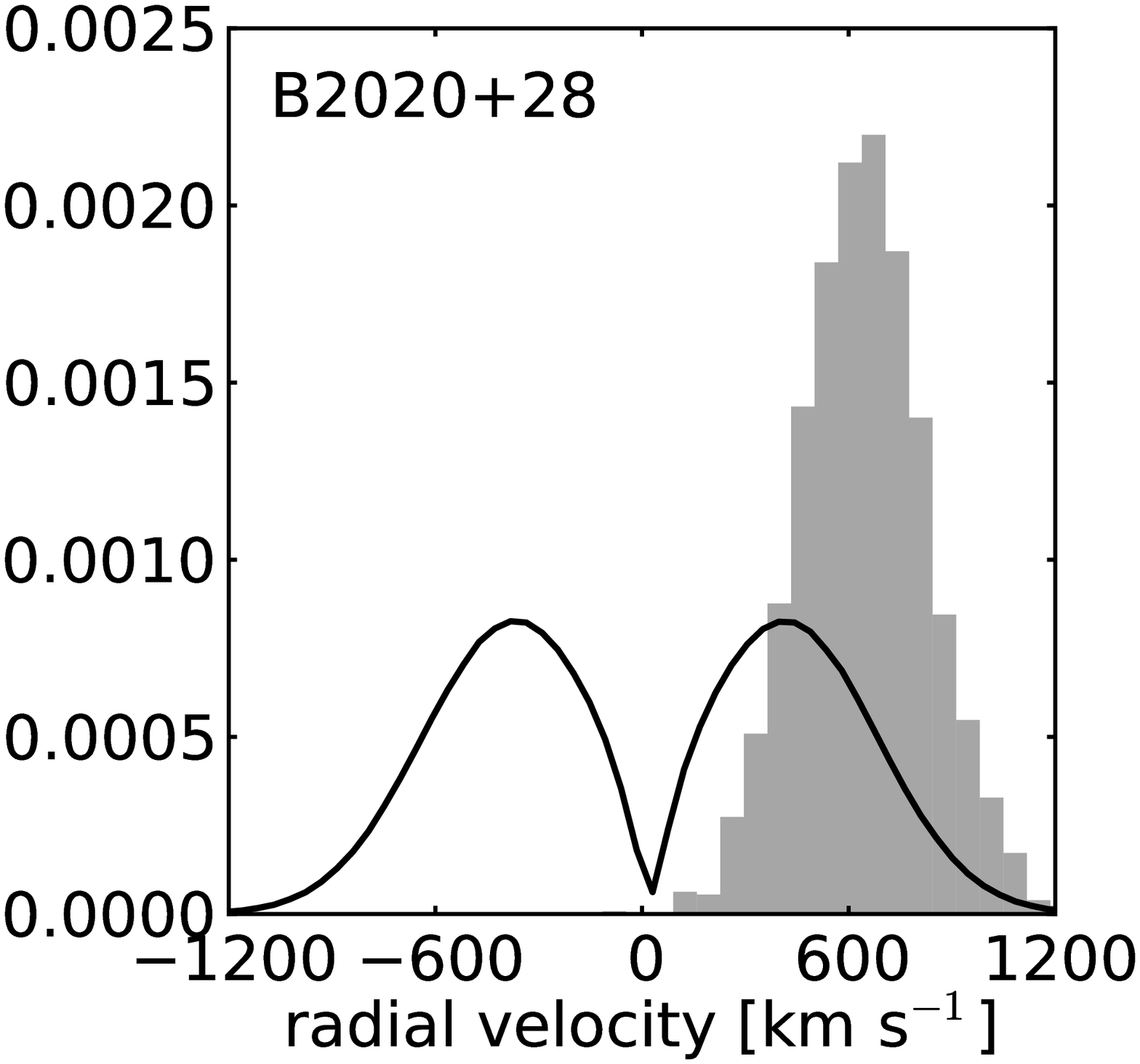}}
\par\end{centering}

\begin{centering}
\vspace{-0.5cm}\subfloat{\includegraphics[width=0.25\textwidth]{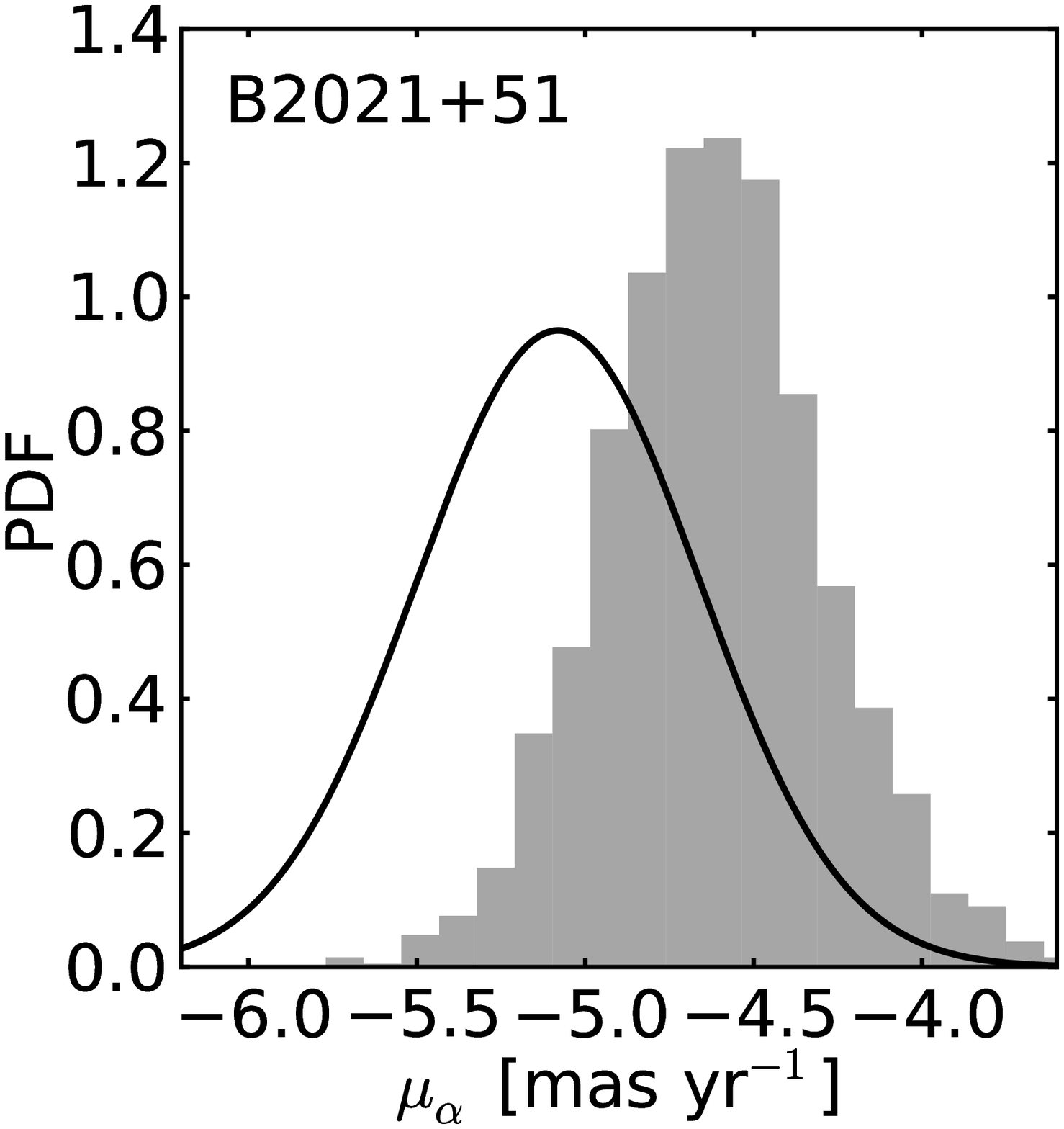}}\subfloat{\includegraphics[width=0.25\textwidth]{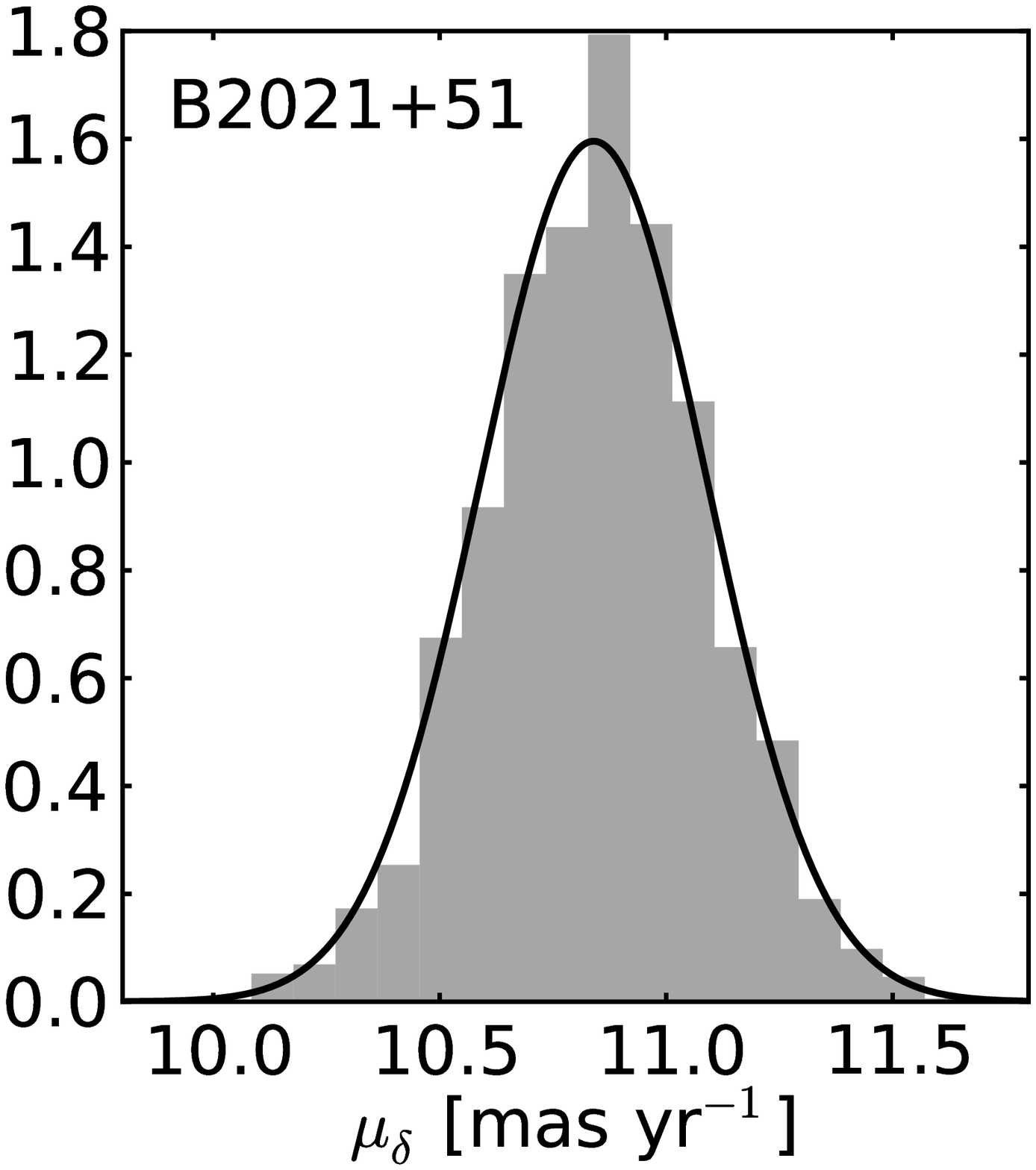}}\subfloat{\includegraphics[width=0.25\textwidth]{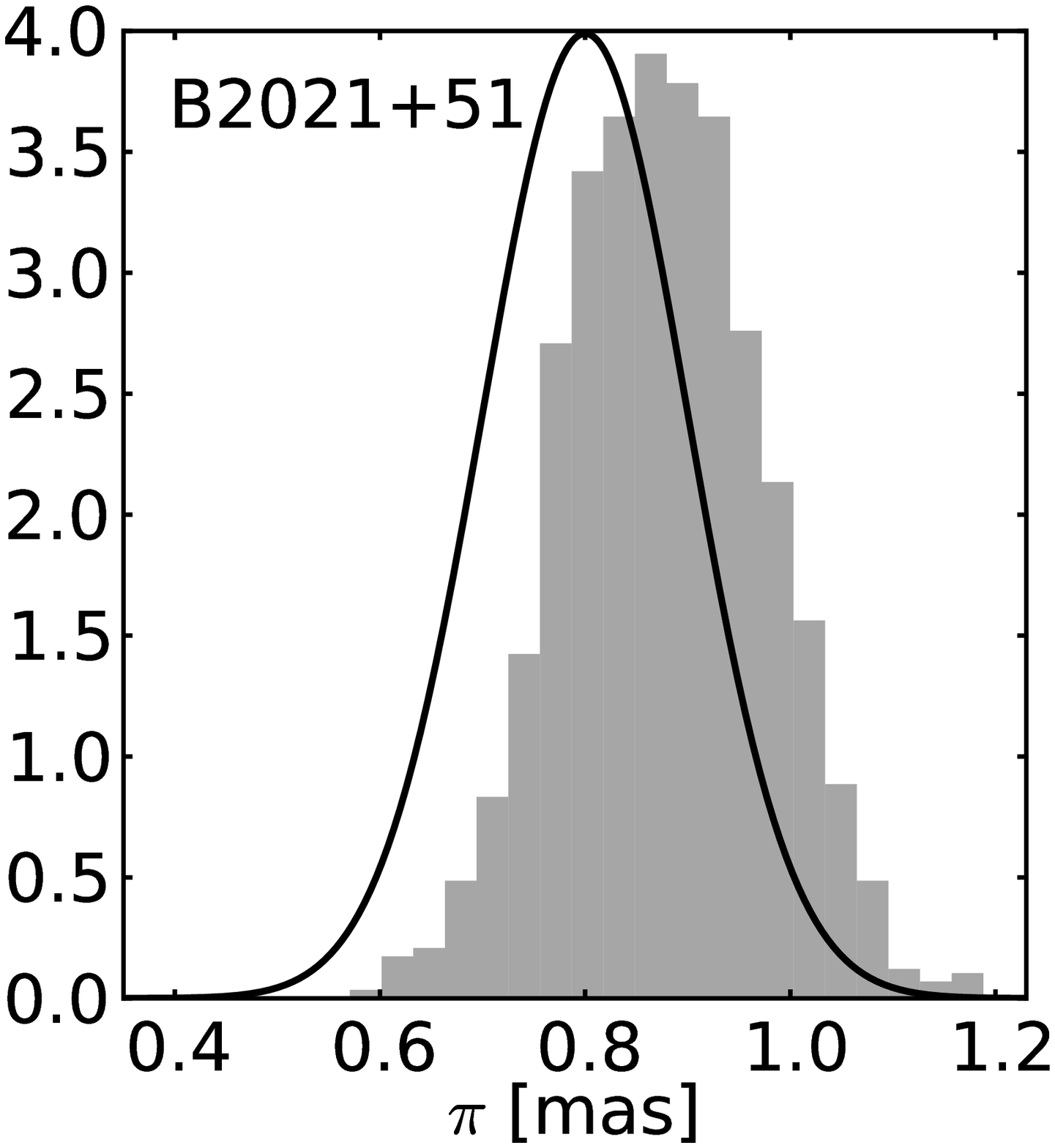}}\subfloat{\includegraphics[width=0.25\textwidth]{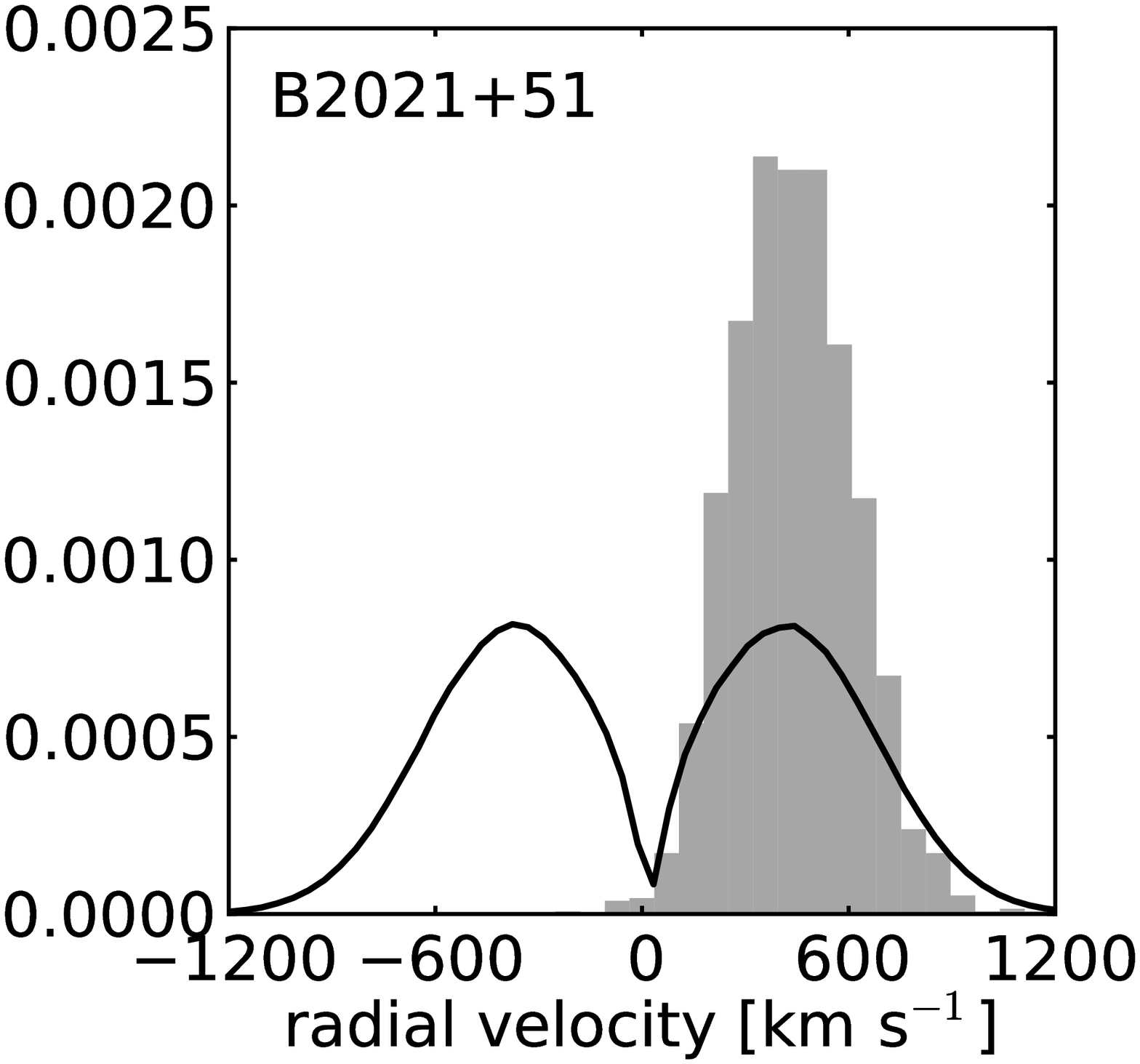}}
\par\end{centering}

\caption[Probability density function of astrometric parameters and required
radial velocities that result in a minimum separation of less than
$10\,$pc between B1929+10/$\zeta$ Oph and B2020+28/B2021+51]{\label{fig:Simulation-histos-1}Probability density function (grey
histograms) of astrometric parameters and required radial velocities
that result in a minimum separation of less than $10\,$pc between
B1929+10/$\zeta$ Oph and B2020+28/B2021+51. Columns from left to
right show the results for $\mu_{\alpha},\,\mu_{\delta},\,\pi,\,$and
$V_{\text{rad}}$. For the measured parameters $\mu_{\alpha},\,\mu_{\delta},\,\text{and }\pi$,
the solid lines indicate the input parameter distributions derived
from assuming Gaussian errors. The solid line in the last column
indicates the input distribution for $V_{\text{rad}}$ as derived
from our measured transverse velocity and the empirically determined
space velocity distribution derived by \citet{hobbs05}. Objects from
top to bottom are B1929+10, $\zeta$ Oph, B2020+28, and B2021+51. }
\end{figure*}

To test how much the updated solar parameters and the different radial
velocity distributions%
\footnote{We used a one-component velocity distribution, while \citet{vlemmings04}
used a two-component distribution.%
} influence the computed trajectories, we also ran the simulations
for B2020+28/B2021+51 using the input parameters for $\mu_{\alpha},\,\mu_{\delta},\,\text{and\ \ensuremath{\pi}}$
from \citet{vlemmings04} (Table \ref{tab:previous-fitting-results}).
In our simulations $0.14\%$ of trajectories cross within $10\,$pc\textcolor{red}{{}
}(minimal distance of $0.10\,$pc), reproducing the results of\textcolor{red}{{}
}these earlier simulations well.\\
\\
In the three million simulations that we ran using our bootstrapping results
for B1929+10 and the latest astrometric parameters for $\zeta$ Oph,
258272 (8.6\%) orbits cross within $10\,$pc about $0.5\,$Myr ago
(Fig. \ref{fig:distance-time-histos}). However, none of these orbits
yield a minimum separation of less than $2.4\,$pc, and neither the
pulsar nor the runaway star approach the centre of Upper Scorpius
to within less than $17\,$pc. The median radial velocity of B1929+10
required for it to approach $\zeta$ Oph within $10\,$pc is $570_{-63}^{+53}\,$km$\,$s$^{-1}$
(Fig. \ref{fig:Simulation-histos-1}). For completeness, we also
tested the hypothesis that B1929+10 once formed a binary system with
the pulsar B1952+29 \citep{wright79}. For the latter we assumed the
proper motion from \citet{hobbs04} and the parallax from the distance
derived from the dispersion measure (DM) in the ATNF Pulsar Catalogue%
\footnote{http://www.atnf.csiro.au/research/pulsar/psrcat/%
} and the \foreignlanguage{british}{Galactic electron density model
from \citealt{cordes03}} (Table \ref{tab:sim-parameters}). We assumed
a parallax uncertainty of $1\,$mas in lieu of a formal error estimate
in the DM-based distance. With these parameters, none of the simulated
orbits crosses within $10\,$pc. \foreignlanguage{british}{The same
is true for simulations ran with the same proper motion parameters,
but with the distance estimate $d=0.42\,$kpc, based on the same DM
but using instead the Galactic electron density model from \citet{taylor93b}.}\\
\\
For the putative pulsar pair B2020+28/B2021+51 our new measurements
imply a minimum possible separation of $1.9\,$pc. Of the three million
trajectories, $1866$ (0.06\%) cross within $10\,$pc about $1.16_{-0.17}^{+0.18}\,$Myr
ago (Fig. \ref{fig:distance-time-histos}). The implied median radial
velocities are $643_{-168}^{+193}$ for B2020+28 and $433_{-193}^{+154}\,$km$\,$s$^{-1}$
 for B2021+51 (Fig. \ref{fig:Simulation-histos-1}).

\section{Discussion\label{sec:Discussion}}

\subsection{Binary companion of B1929+10}

Our new astrometric results for the pulsar B1929+10 confirm the measurements
of earlier VLBI campaigns \citep{brisken02,chatterjee04}, and in combination
with the previous position measurements, we place robust constraints
on the uncertainties of the proper motion parameters. Accordingly,
for our adopted parallax $\pi=2.77_{-0.08}^{+0.07}$ our values for
the proper motion differ by more than $10\sigma$ from the parameter
space that implies a common origin of B1929+10 and $\zeta$ Oph in
Upper Scorpius in \citet{hoogerwerf01}. Given the new astrometry
for the pulsar and also the updated astrometric parameters for the
runaway star, the minimal possible separation of $2.4\,$pc between
the two stars is too large to be consistent with a common origin of
both. Moreover, the closest approach of roughly $17\,$pc to Upper
Scorpius of either of the two objects in all of the trajectories crossing
within $10\,$pc makes it very unlikely that this region is the place of
common origin. \\
\\
However, the fraction of simulated orbits that cross within $10\,$pc
($\sim8.6\%$) is surprisingly high and implies that the orbits of
both objects may have crossed within that distance about $0.5\,$Myr
ago. The allowed range in radial velocities for them to pass close
by, on the other hand, is very small and points to a very strong kick
imparted to the pulsar at birth. Only a direct measurement of the
pulsar's radial velocity will further constrain the distance of closest
approach of B1929+10 and $\zeta$ Oph.\\
\\
Our data in combination with the updated astrometry for B1952+29 make
a binary origin of B1929+10 and B1952+29 highly implausible.

\subsection{Common origin of B2020+28 and B2021+51}

\citet{vlemmings04} used the astrometric parameters from \citet{brisken02}
(Table \ref{tab:previous-fitting-results}) to infer a putative common
origin of B2020+28 and B2021+51. This conclusions seems plausible
considering that the pulsars' 2D motions
lie in apparently opposite directions in Galactic coordinates (Fig.
1 in \citealt{vlemmings04}) and also because the pulsars have very similar
characteristic ages of $2.88$ (B2020+28) and $2.75\,$Myr (B2021+51). Nevertheless, our new proper motion
measurements, in conjunction with our parallax measurements, which
place both pulsars almost twice as close as previous distance estimates,
rule out a common origin for these two objects. \citet{vlemmings04}
determined the percentage of orbits crossing within $10\,$pc for
a known binary disrupted $1\,$Myr ago as a function of astrometric
uncertainties (see their Fig. 2). These models indicate that our
improved errors should have yielded $1\%$ of crossing orbit realization
(within $10\,$pc) for B2020+28 and B2021+51. However,
in our simulations only $0.06\%$ of trajectories cross within that
distance, and none of the orbits yield an approach of less than $1.9\,$pc.
Even if we use the bootstrapping results with their larger errors
from Table \ref{tab:my-fitting-results}, only $0.08\%$ of the
orbits
cross within $10\,$pc. Furthermore, the orbits approaching each other
within $10\,$pc do so at a median distance of $0.64_{-0.11}^{+0.09}\,$kpc
to the solar system. Given the estimated extent of the Cygnus Superbubble
of $0.7-2.5\,$kpc \citep{vlemmings04}, a common origin within this
region is ruled out.

\subsection{Implications for Galactic electron density along the lines of sight}

\begin{table}[!t]
\caption[\selectlanguage{british}%
Astrometry used in simulations\selectlanguage{english}%
]{\label{tab:dm-distances}\foreignlanguage{british}{}Parallax- vs.
DM-based distances }

\begin{centering}
\begin{tabular}{lcccc}
\noalign{\vskip-0.3cm}
\hline\hline &  &  &  & \tabularnewline
\noalign{\vskip-0.2cm}
 & DM & $d_{\pi}$\tablefootmark{a} & $d_{{\rm NE2001}}$\tablefootmark{b} & $d_{{\rm TC93}}$\tablefootmark{c}\tabularnewline
Source & {[}pc$\,$cm$^{-3}${]} & {[}kpc{]} & {[}kpc{]} & {[}kpc{]}\tabularnewline
\hline 
\noalign{\vskip\doublerulesep}
B1929+10 & $3.180\,(4)$ & $0.361_{-0.010}^{+0.009}$ & $0.34$ & $0.17$\tabularnewline
B2020+28 & $24.640\,(3)$ & $1.39_{-0.06}^{+0.05}$ & $2.11$ & $1.30$\tabularnewline
B2021+51 & $22.648\,(6)$ & $1.25_{-0.17}^{+0.14}$ & $1.94$ & $1.22$\tabularnewline
\hline\vspace{-0.4cm} &  &  &  & \tabularnewline
\end{tabular}
\par\end{centering}

\tablefoot{
\tablefoottext{a}{Parallax-based distances derived in this work.}
\tablefoottext{b}{Distance estimate based on the Galactic electron density model from Cordes and Lazio (2003)}
\tablefoottext{c}{Distance estimate based on the Galactic electron density model from Taylor et al. (1993).}\\
Except for $d_\pi$ , all values were taken from the ATNF pulsar catalogue.
}
\end{table}
In Table \ref{tab:dm-distances} we list the distances inferred from
our parallax measurements in comparison with those implied by the
DM and the Galactic electron density models of \citet[NE2001]{cordes03}
and \citet[TC93, both obtained from the ATNF pulsar catalogue]{taylor93}.
While the NE2001--distance agrees with our measurement for
B1929+10, the same model overestimates the distances to both B2020+28
and B2021+51 by a factor of about $1.5$. The distance estimates for
the latter two pulsars as given by the preceding model, TC93,
agree well with our results, however. Hence, in combination with
the DM as listed in Table \ref{tab:dm-distances}, our parallax measurements
imply a mean electron density of $8.8_{-0.2}^{+0.3},\,17.7_{-0.6}^{+0.8},\,{\rm and}\,18.1_{-1.8}^{+2.8}\,{\rm cm}^{-3}$
along the lines of sight to B1929+10, B2020+28, and B2021+51, respectively.

\section{Conclusions\label{sec:Conclusions}}

Based on our new astrometry for pulsars B1929+10, B2020+28, and B2021+51
obtained with the EVN at 5$\,$GHz, we rule out previously proposed
common origin scenarios for all three sources. Our Monte Carlo simulations
of the past trajectory of B1929+10 throughout the Galactic potential
show now indication for the pulsar to have once been in a binary system
with the runaway star $\zeta$ Oph in Upper Scorpius. Similar simulations
for B2020+28 and B2021+51 also rule out a binary origin of the pulsars
in the Cygnus Superbubble.
\begin{acknowledgements}
We appreciate the comments of the anonymous referee that
helped
us to improve the manuscript. We would like to thank Walter Brisken
for providing us with his position measurements at 1.5$\,$GHz. F.K.
acknowledges partial funding by the Bonn Cologne Graduate School of
Physics and Astronomy. The European VLBI Network is a joint facility
of European, Chinese, South African and other radio astronomy institutes
funded by their national research councils. This work has been supported
by the European Commission Framework Programme 7, Advanced Radio Astronomy
in Europe, grant agreement No. 227290.
\end{acknowledgements}
\bibliographystyle{aa}
\bibliography{biblio}

\end{document}